\documentclass[aps,twocolumn,superscriptaddress,pre]{revtex4-2}

\usepackage{bm,graphicx,textcomp,amssymb,amsmath,dcolumn,color}
\usepackage{hyperref}
\usepackage{tabularx}
\usepackage{enumitem}
   
\usepackage[utf8]{inputenc}
\usepackage[T1]{fontenc}

\usepackage{notes2bib}
\newcommand{\ket}[1]{\left|#1\right>}
\newcommand{\bra}[1]{\left<#1\right|}

\newcommand{\nn}{\nonumber\\}

\newcommand{\f}[1]{\mbox{\boldmath$#1$}}

\newcommand{\bea}{\begin{eqnarray}}
\newcommand{\ea}{\end{eqnarray}}
\newcommand{\eea}{\end{eqnarray}}

\newcommand{\ii}{{\rm i}}
\newcommand{\abs}[1]{{\left| #1 \right|}}
\newcommand{\expval}[1]{{\langle #1 \rangle}}
\newcommand{\trace}[1]{{\rm Tr}\left\{#1\right\}}
\newcommand{\traceS}[1]{{\rm Tr_S}\left\{#1\right\}}
\newcommand{\traceB}[1]{{\rm Tr_B}\left\{#1\right\}}

\newcommand{\new}[1]{{#1}}%{{\color{red} #1}}

\begin{document}

\title{Finite-time performance of a cyclic 2d quantum Ising heat engine}

\author{S. P. Katoorani}
\affiliation{Helmholtz-Zentrum Dresden-Rossendorf, Bautzner Landstra{\ss}e 400, 01328 Dresden, Germany}

\author{C. Kohlf\"urst}
\affiliation{Helmholtz-Zentrum Dresden-Rossendorf, Bautzner Landstra{\ss}e 400, 01328 Dresden, Germany}

\author{F. Queisser}
\affiliation{Helmholtz-Zentrum Dresden-Rossendorf, Bautzner Landstra{\ss}e 400, 01328 Dresden, Germany}

\author{G. Schaller}
\email{g.schaller@hzdr.de}
\affiliation{Helmholtz-Zentrum Dresden-Rossendorf, Bautzner Landstra{\ss}e 400, 01328 Dresden, Germany}

\author{R. Sch\"utzhold}
\affiliation{Helmholtz-Zentrum Dresden-Rossendorf, Bautzner Landstra{\ss}e 400, 01328 Dresden, Germany}
\affiliation{Institut f\"ur Theoretische Physik, Technische Universit\"at Dresden, 01062 Dresden, Germany}
\date{\today}

\begin{abstract}
We discuss the limit cycle regime of a finite-time quantum Otto cycle with a frictionless two-dimensional anisotropic Ising model as the working fluid.
From Onsagers exact equilibrium solution, we first find optimal parameters for the operational modes of work extraction and cooling for infinitely slow cycles.
The equilibrium points in these optimal cycles correspond to different phases of the model, such that the non-equilibrium dynamics during the cycle bypasses the phase transition.
Finite-time cycles allow for finite power extraction or cooling currents, but for such cycles we point out that -- already within the regime of weak system-reservoir coupling -- 
energetic changes of the system during dissipative strokes may contain a significant portion of coupling and decoupling control work and should thus not be directly identified with heat.
For ultrafast cycles, the required control work spoils performance, such that to maximize work extraction or cooling heat per cycle time, there is an optimal cycle duration.
We also find that net zero-energy transitions may lead to undesired reservoir heating.
\end{abstract}

\maketitle

%%%%%%%%%%%%%%%%%%%%%%%%%%%%%%%%%%%%%%%%%%%%%%%%%%%%%%%%%%%%%%%%%%%%%%%%%%%%%%%%
%%%%%%%%%%%%%%%%%%%%%%%%%%%%%%%%%%%%%%%%%%%%%%%%%%%%%%%%%%%%%%%%%%%%%%%%%%%%%%%%
\section{Introduction} 
%%%%%%%%%%%%%%%%%%%%%%%%%%%%%%%%%%%%%%%%%%%%%%%%%%%%%%%%%%%%%%%%%%%%%%%%%%%%%%%%
%%%%%%%%%%%%%%%%%%%%%%%%%%%%%%%%%%%%%%%%%%%%%%%%%%%%%%%%%%%%%%%%%%%%%%%%%%%%%%%%

Quantum thermodynamics promises interesting applications at the nanoscale~\cite{binder2019,cangemi2024a}.
While quantum thermodynamic devices are subject to the same Carnot bounds as classical ones,
collective quantum effects could still be used to enhance the performance of quantum heat engines~\cite{hardal2015a,uzdin2016a,jaramillo2016a,niedenzu2018a,kloc2019a,watanabe2020a,macovei2022a,da_silva_souza2022a,kolisnyk2023a}.
Additionally, from a foundational perspective, the study of such devices has triggered the transfer of
fundamental classical notions of work and heat into the quantum domain~\cite{campisi2011a}. 

We can roughly distinguish three classes of quantum heat engines.
First, in continuously operating engines~\cite{kosloff2014a}, the working fluid is simultaneously and at all times coupled to multiple reservoirs
maintained at different thermal equilibrium states.
In absence of chemical potentials, one then needs at least three reservoirs to perform useful functions like e.g. steady-state cooling of the coldest reservoir~\cite{levy2012b}.
Second, one may also consider heat engines that are coupled to (typically two) reservoirs and simultaneously subject to external driving~\cite{scovil1959a,szczygielski2013a,scopa2018a,kalaee2021a,kalaee2021b,kolisnyk2024a}.
The co-existence of dissipative and driven dynamics makes the proper calculation of heat currents and work extraction statistics rather challenging.
The analysis becomes much simpler in the third class of finite-stroke engines~\cite{feldmann2000a,kieu2004a}, where, analogous to classical thermodynamic cycles, the working fluid is alternatingly subject to unitary strokes, during which it evolves unitarily without heat exchange, and dissipative strokes, during which it is coupled to a thermal reservoir.
Ideally, the cycle picture would allow for a straightforward classification of the system energetic changes as work and heat, respectively, and thus simplify the performance assessment.

Therefore, for interacting spin systems that are challenging on their own one often focuses on the class of finite-stroke engines, for which many interacting quantum working fluids have already been analyzed.
One may object that the actual demonstration of work extraction in this class typically requires a flywheel analogue that is energetically charged during the operation~\cite{marchegiani2016a,seah2018a,strasberg2021a,martins2023a}, but also other operational modes e.g. of refrigeration by investing work can be considered.
Very often, one considers the infinitely slow regime of quantum-adiabatic unitary transformations and infinitely long dissipative strokes.
While important for qualitative analysis, this limit is practically not so relevant as the power (work per cycle time) and cooling power (cooling heat per cycle time) vanish
for infinite cycle times.
Therefore, researchers have attempted to discuss finite-time cyclic operations with various quantum working fluids~\cite{feldmann2000a,wang2012c,kloc2019a,das2020a}.
Working fluids with a phase transition are particularly interesting in this respect, as they promise near Carnot efficiency at finite power~\cite{campisi2016a}.
For quantum-critical working fluids, the closure of the energy gap during the unitary strokes has been identified as an obstacle to fast and quantum adiabatic cycle operation~\cite{alecce2015a,revathy2020a}, but also remedies like e.g. shortcuts to adiabaticity~\cite{delcampo2014a,beau2016a,cakmak2019a,hartmann2020b} have been proposed.
However, finite-stroke quantum heat engines need not be operated quantum-adiabatically.
To realize a stroke without heat transfer, any unitary closed evolution of the quantum working fluid 
is formally sufficient.
In particular, working fluids driven by a Hamiltonian that commutes with itself at different times~\new{\cite{niedenzu2018a,dann2020a,mukherjee2020a,latune2023a}} will not suffer from
a rotation of eigenvectors, which allows for an intrinsically frictionless evolution:
States that are initially diagonal in the system energy eigenbasis will remain constant independent of the driving speed, such that the unitary strokes can be reduced to instantaneous quenches without altering the cycle performance.
Consequently, for such models, the dissipative strokes limit the cycle time.
For these however, the effect of the reservoir can no longer be described by Fermi-Golden-Rule rates in the fast cycle regime, and non-Markovian effects have to be taken into account~\cite{wiedmann2020a,ptaszynski2022a,ishizaki2023a,picatoste2024a}.
For example, for extremely short coupling times the quantum Zeno effect~\cite{itano1990a} may freeze relaxation processes, and it has been argued that it may eventually limit
the cycle efficiency~\cite{mukherjee2020a,shirai2021a,picatoste2024a}.

So far, these extreme scenarios have been investigated for non-interacting simpler systems, that however have met experimental implementations on several platforms~\cite{rossnagel2016a,maslennikov2019a,klatzow2019a,peterson2019a,lisboa2022a,zhang2022a,koch2023a,uusnaekki2025a,shende2025a}.
In the present contribution we aim to study these effects in an interacting spin system and analyze a finite-time quantum Otto cycle with a two-dimensional quantum Ising model subject to a frictionless drive.
We start by reviewing the quantum Otto cycle in Sec.~\ref{SEC:FTQO} and discuss the associated finite-time transition rates as well as required definitions for heat and work.
To gain some intuition we first consider a two-level system as working fluid example in Sec.~\ref{SEC:tls}, for which we obtain mostly analytic results also in the limit cycle regime.
Afterwards, we discuss our results for the anisotropic Ising model in Sec.~\ref{SEC:a2dim} before concluding.
Technical derivations can be found in the appendices.

%%%%%%%%%%%%%%%%%%%%%%%%%%%%%%%%%%%%%%%%%%%%%%%%%%%%%%%%%%%%%%%%%%%%%%%%%%%%%%%%
%%%%%%%%%%%%%%%%%%%%%%%%%%%%%%%%%%%%%%%%%%%%%%%%%%%%%%%%%%%%%%%%%%%%%%%%%%%%%%%%
\section{Finite-time quantum Otto cycle}\label{SEC:FTQO} 
%%%%%%%%%%%%%%%%%%%%%%%%%%%%%%%%%%%%%%%%%%%%%%%%%%%%%%%%%%%%%%%%%%%%%%%%%%%%%%%%
%%%%%%%%%%%%%%%%%%%%%%%%%%%%%%%%%%%%%%%%%%%%%%%%%%%%%%%%%%%%%%%%%%%%%%%%%%%%%%%%

\subsection{Quantum Otto cycle}

\begin{figure}
\includegraphics[width=0.4\textwidth]{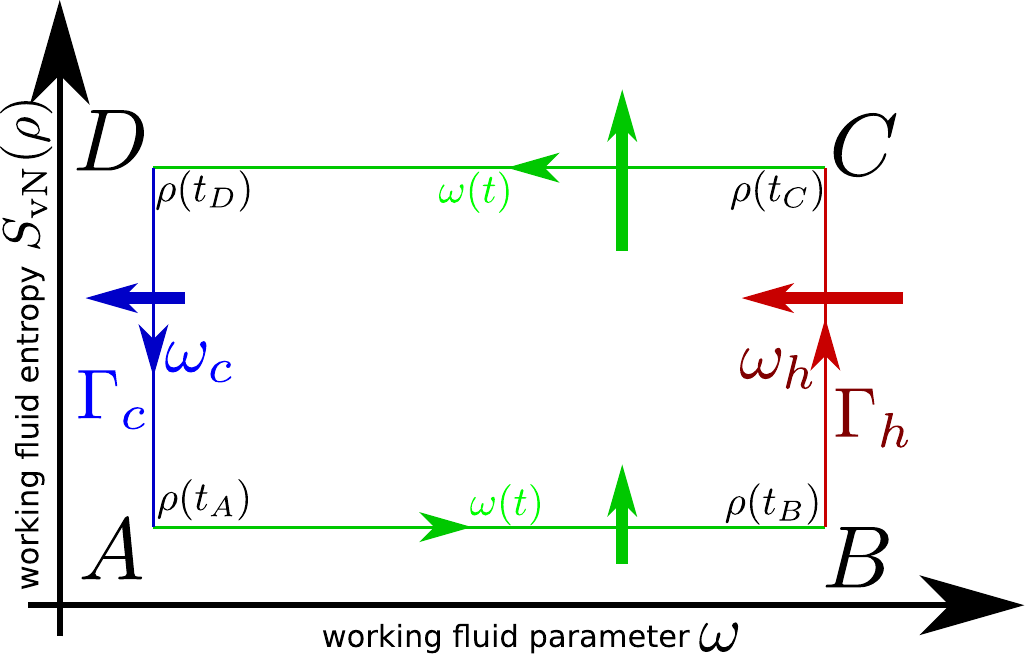}
\caption{\label{FIG:quantumotto}
Sketch of a quantum Otto cycle in the plane of the working fluid quench parameter $\omega$ and the von-Neumann entropy of the working fluid, cyclically operated via $A\to B\to C\to D\to A$.
The Hamiltonian is constant while coupled to the cold ($D\to A$, blue) or to the hot ($B\to C$, red) reservoir.
During the isentropic strokes (green), the system is subject to a time-dependent system Hamiltonian but decoupled from both reservoirs.
Bold arrows exemplarily sketch the flow of energy into (inward-pointing) or out of (outward-pointing) the working fluid for work extraction: 
\new{Positive heat from the hot reservoir increases the working fluid energy (red arrow), whereas the energy balance with the cold reservoir is negative (blue arrow).
Ideally, the difference of the heat contributions (length of red and blue arrows) is available for the net extraction
of work (length of green arrows).
For the refrigerator operational mode the heat and work flows need to be inverted.}
}
\end{figure}

In the classical Otto cycle, one has two isentropic strokes (only work) and two isochoric ones (only heat exchange).
A quantum analog cycle that (at least in principle) allows for a similarly simple distinction between work and heat 
is known as quantum Otto cycle~\cite{kosloff2017a}.
It is composed of four strokes
(see Fig.~\ref{FIG:quantumotto} for an illustration):
\begin{itemize}[leftmargin=0.07\textwidth]
\item[$A\to B$]During this isentropic stroke, the working fluid only evolves under the time-dependent Hamiltonian $H_S(t)$ such that $H_S(t_A) = H_S^c$ and $H_S(t_B) = H_S^h$.
Accordingly, its entropy remains constant, its energetic changes are interpreted as work, and the reservoirs need not be explicitly modeled.
\item[$B\to C$]During this dissipative stroke (the analog of the isochoric stroke in the classical Otto cycle), the working fluid is subject to the constant system Hamiltonian $H_S^h$ and also to the coupling $H_I^h$ to the hot reservoir $H_B^h$. Energetic changes of the working fluid are typically interpreted as heat.
\item[$C\to D$]This isentropic stroke is analogous to the stroke $A\to B$, except that the transform is reversed, i.e., $H_S(t_C) = H_S^h$ and $H_S(t_D) = H_S^c$.
\item[$D\to A$]The cycle is closed by this dissipative stroke, during which the working fluid is subject to the constant system Hamiltonian $H_S^c$ and the coupling $H_I^c$ to the cold reservoir $H_B^c$.
\end{itemize}
After completion of one cycle, it is restarted with the stroke $A\to B$.
At any time, the total Hamiltonian of the model universe is thus comprised of 
\begin{align}
H(t) = H_S(t) + H_B^c + H_B^h + H_I^c(t) + H_I^h(t)\,,
\end{align}
where during every stroke of one cycle, different parts of the Hamiltonian are active.

For the heat engine operational mode, it is desirable to maximize the work extracted during one \new{cycle $\Delta W$, which is by our conventions positive when work is performed by the working fluid.
For a closed cycle, energy conservation dictates that $\Delta W = \Delta Q_h + \Delta Q_c$, where $\Delta Q_{h/c}$ denote the heat entering the working fluid (i.e., positive when increasing its energy) during the respective dissipative strokes.}
Thus, when we interpret all system energetic changes during the dissipative strokes as the corresponding negative changes in the reservoir (i.e., as heat), 
this allows us to estimate the extracted work from the difference of system energies
\begin{align}\label{EQ:worksys}
\Delta W_{\rm sys} = -(E_B-E_A)-(E_D-E_C)\,.
\end{align}
For the refrigerator operational mode one would like to maximize the heat leaving the cold reservoir, which under the same identification is estimated by the energy
inflow into the system during that stroke
\begin{align}\label{EQ:heatsys}
\Delta Q_{c, \rm sys} = E_A-E_D\,.
\end{align}
These quantities can e.g. be used to define the efficiency for the heat engine operational mode
$\eta = \frac{\Delta W}{\Delta Q_h} \Theta(\Delta W) \le 1 - \frac{\beta_h}{\beta_c} = \eta_{\rm Ca}$
and the coefficient of performance for the refrigerator operational mode
$\kappa = \frac{\Delta Q_c}{-\Delta W} \Theta(\Delta Q_c) \le \frac{\beta_h}{\beta_c-\beta_h} = \kappa_{\textrm Ca}$
which are both bound by their respective Carnot limits.
Analogous relations would hold for the heat pump operational mode aiming at heating the hot reservoir, which we will not discuss here.

\subsection{Finite-time transition rates}

Very often, Fermi-Golden-Rule (FGR) rates are used to model relaxation processes. 
In their derivation, one assumes that reservoir and system are coupled for an infinitely long time.
For example, when the system is coupled to reservoir $\nu$ that is composed of a continuum of two-level systems held at thermal equilibrium with inverse temperature $\beta_\nu$, 
the rate for the system undergoing an energy change of $+\Omega$ can in the FGR limit (see App.~\ref{APP:coarsegraining}) be expressed as
\begin{align}\label{EQ:rates_fgr}
R^{\nu,\rm FGR}_\Omega = \frac{\Gamma_\nu(\Omega)}{e^{\beta_\nu \Omega}+1}\,,
\end{align}
where the reservoir spectral function $\Gamma_\nu(\omega)\ge 0$ is analytically continued to the real axis via $\Gamma_\nu(\omega)=\Gamma_\nu(-\omega)$.
For this paper, we will use a Lorentzian representation
\begin{align}\label{EQ:spectral_function}
\Gamma_\nu(\omega) = \frac{\Gamma_\nu \delta_\nu^2}{\omega^2+\delta_\nu^2}\,,
\end{align}
where for simplicity we constrain ourselves to symmetric coupling strengths $\Gamma_\nu=\Gamma$ and widths $\delta_\nu=\delta$ for both reservoirs.
The FGR rates above obey detailed balance $R^{\nu,\rm FGR}_{-\Omega} = e^{\beta_\nu\Omega} R^{\nu,\rm FGR}_{+\Omega}$, 
such that relaxation processes favor low-energy states and the stationary state is the thermal equilibrium state of the system at inverse temperature $\beta_\nu$.

For very short stroke durations one cannot expect the FGR treatment to apply.  
Rather, second order perturbation theory allows to derive effective rates that are coarse-grained over the stroke duration $\tau$~\cite{schaller2008a}.
\new{
While by construction, the coarse-grained rates only yield trustworthy results at the end of the stroke, the rate-equation representation allows to apply standard trajectory techniques (see App.~\ref{APP:stochprop}) also to non-Markovian settings.
}
Specifically, the rate for an energy change $\Omega$ observed in the system is then (see App.~\ref{APP:coarsegraining})
\begin{align}\label{EQ:ratecg}
R^{\nu,\tau}_\Omega = \int \frac{\Gamma_\nu(\omega)}{e^{\beta_\nu \omega}+1} \frac{\tau}{2\pi} {\rm sinc}^2\left[\frac{(\omega-\Omega)\tau}{2}\right]d\omega\,,
\end{align}
where ${\rm sinc}(x)\equiv \sin(x)/x$.
These coarse-grained rates do in general not obey detailed balance but fall back to the FGR rates~\eqref{EQ:rates_fgr} when $\tau\to\infty$.
Furthermore, in the wideband limit ($\delta_\nu\to\infty$, i.e., $\Gamma_\nu(\omega)\to \Gamma_\nu$), the coarse-grained rates still favor energetically lower states at all $\tau$.
In the limit $\tau\to 0$ (and supposing finite $\delta_\nu$) all rates will vanish, such that for repeated short couplings to a reservoir the system will not undergo any transitions anymore.
The repeated coupling with an uncorrelated reservoir is analogous to repeated measurements, such that this inhibition of transitions by fast repeated couplings can be seen as a manifestation of the quantum Zeno effect~\cite{itano1990a,gurvitz2003a,ahmadiniaz2022a}.
That rates vanish for small $\tau$ also means that cycles can be expected to become dysfunctional when driven too fast, but note that spectral functions with more structure than~\eqref{EQ:spectral_function} can be exploited to enhance cycle performance for finite $\tau$ via the Anti-Zeno effect~\cite{kofmann2000a,mukherjee2020a}.

When the system suffers a transition with energy change $\Omega$, the corresponding average energy change in the reservoir $\nu$ is (see App.~\ref{APP:coarsegraining})
\begin{align}\label{EQ:reschangecg}
\Delta E^{\nu,\tau}_\Omega = -\frac{\int \frac{\omega \Gamma_\nu(\omega)}{e^{\beta_\nu\omega}+1} \frac{\tau}{2\pi} {\rm sinc}^2\left[\frac{(\omega-\Omega)\tau}{2}\right]d\omega}{R^{\nu,\tau}_\Omega}\,.
\end{align}
From this it follows that in general $\Delta E^{\nu,\tau}_\Omega \neq -\Omega$, i.e., for finite stroke durations there will be a mismatch between system and reservoir energy changes.
This mismatch is invested in the control work used for coupling and decoupling to the reservoir~\cite{schaller2020a} and should enter the thermodynamic discussion of finite-time cycles, see below.
For infinitely slow cycles we find that $\lim_{\tau\to\infty} \Delta E^{\nu,\tau}_\Omega = -\Omega$.
In the particular case of energy-preserving system transitions ($\Omega=0$), one finds from integral transformations that $R^{\nu,\tau}_0 \Delta E^{\nu,\tau}_0 \ge 0$, which means that these transitions
will only induce indiscriminate reservoir heating, detrimental for cycle performance.
A similar statement is found for sequential transitions of opposite energetic changes, where one formally finds
$R^{\nu,\tau}_{+\Omega} \Delta E^{\nu,\tau}_{+\Omega} + R^{\nu,\tau}_{-\Omega} \Delta E^{\nu,\tau}_{-\Omega} \ge 0$.
This means that these processes will on average also just induce heating of the reservoir without net energy change of the system and should be detrimental to cycle performance.

\subsection{Finite-time definitions of heat and work}

As exposed above, for finite contact times $\tau$, the change of the cold and hot reservoir energies 
$\Delta E^{c,\tau}_{D\to A}$ and $\Delta E^{h,\tau}_{B\to C}$ is not just the negative of the system energy changes -- even in the weak-coupling regime.
We therefore define the control work required for coupling and decoupling the reservoirs as the difference
\begin{align}\label{EQ:controlwork}
-\Delta W_{\rm ctl}^\tau &= \Delta E^c_{D\to A} + E_A-E_D\nn
&\qquad	 + \Delta E^h_{B\to C} + E_C-E_B\,,
\end{align}
such that the net work extracted from the cycle is 
\begin{align}\label{EQ:network}
-\Delta W_{\rm net}^\tau &= -\Delta W_{\rm ctl}^\tau - \Delta W_{\rm sys}\nn
&= \Delta E^{c,\tau}_{D\to A} + \Delta E^{h,\tau}_{B\to C}\,,
\end{align}
which requires to calculate the reservoir energetic changes.
Also the net heat extracted from the cold reservoir is given by the corresponding energy change of the reservoir, e.g. for cooling 
\begin{align}\label{EQ:netheat}
\Delta Q_c^\tau &= -\Delta E^{c,\tau}_{D\to A}\,.
\end{align}

To evaluate these contributions, let us highlight two crucial assumptions:
\begin{itemize}
\item 
First, we assume that the evolution is frictionless as far as the working fluid is concerned
\begin{align}
\left[H_S(t),H_S(t')\right]=0\,,
\end{align}
which implies that during the isentropic strokes, the time evolution operator for the system can be written without time-ordering
\begin{align}
U_S^{A/C\to B/D} &= \exp\left\{-\ii\int\limits_{t_{A/C}}^{t_{B/D}} H_S(t') dt'\right\}\,,
%\nnU_S(C\to D) &= \exp\left\{-\ii\int_{t_C}^{t_D} H_S(t') dt'\right\}\,,
\end{align}
and that there is a time-independent energy eigenbasis diagonalizing the system Hamiltonian $H_S(t)=\sum_a E_a(t) \ket{a}\bra{a}$.

\item 
Second, we assume the system density matrix to be diagonal (though not necessarily thermal) in the time-independent energy eigenbasis of $H_S(t)$ at all points $A$, $B$, $C$, and $D$, which beyond an FGR treatment is not a trivial simplification.
The above time evolution operator already implies that diagonal density matrices at points $A$ and $C$ will also remain diagonal at points $B$ and $D$, respectively.
In App.~\ref{APP:general_derivation} we detail that under locality assumptions on the system-reservoir couplings, diagonality of the density matrix in the system energy eigenbasis
is also maintained during the dissipative strokes $D\to A$ and $B\to C$.
Therefore, this assumption can be fulfilled by initializing the cycle in a diagonal state, e.g. by using a sufficiently long pre-equilibration with the cold reservoir before
starting the cyclic operations.

Alternatively, one may also imagine that further dephasing reservoirs (whose energy balance is not part of our analysis) active during the dissipative strokes may erase the coherences without changing the systems populations. 
\end{itemize}
Under these assumptions, the system state will not suffer any change during the isentropic strokes.
Thus, it is advisable to perform these strokes infinitely fast, 
and the remaining bottleneck to power extraction or cooling power is posed by the dissipative stroke durations.

%%%%%%%%%%%%%%%%%%%%%%%%%%%%%%%%%%%%%%%%%%%%%%%%%%%%%%%%%%%%%%%%%%%%%%%%%%%%%%%%
%%%%%%%%%%%%%%%%%%%%%%%%%%%%%%%%%%%%%%%%%%%%%%%%%%%%%%%%%%%%%%%%%%%%%%%%%%%%%%%%
\section{Two-level system}\label{SEC:tls}

As an example, we study a two-level system~\cite{kieu2004a}
\begin{align}
H_S(t) = \frac{\omega(t)}{2} \sigma^z\,,
\end{align}
where $\omega(t)=\omega_c$ during $D\to A$ and $\omega(t) = \omega_h$ during $B \to C$.
The coupling to the reservoirs can be described by $H_I^c = \sigma^x \otimes B^c(t)$ and $H_I^h = \sigma^x \otimes B^h(t)$, 
where the reservoir coupling operators $B^c(t)$ and $B^h(t)$ are only non-vanishing during the dissipative cold ($D\to A$) and hot ($B\to C$) strokes of the cycle, respectively.
%Ist "FRG treatment" wirklich ein etablierter Begriff? Ich muss da an "functional renormalization group" denken ==> FGR?
Within a Fermi-Golden-Rule (FGR) treatment, their explicit form will determine the transition rates between ground state $g$ and excited state $e$.
For reservoirs $\nu\in\{c,h\}$ we consider the Glauber rates~\eqref{EQ:rates_fgr} evaluated at energy difference $\Omega\in\pm \omega_\nu$ and also symmetrically coupled wideband reservoirs
$\Gamma_\nu(\omega) = \Gamma$, i.e., 
\begin{align}
R^\nu_{g\to e} = R^{\nu,\rm FGR}_{+\omega_\nu}\,,\qquad
R^\nu_{e\to g} = R^{\nu,\rm FGR}_{-\omega_\nu} = e^{+\beta_\nu\omega_\nu} R^\nu_{g\to e}\,.
\end{align}
For simplicity, we also choose the stroke duration for both dissipative strokes equal and denote it by $\tau$, such that the total cycle duration is $2\tau$. 
This allows for a mostly analytic treatment that we expose in App.~\ref{APP:two-level}.

%"First,... Second,... Third..." auf Seite 3.Liest sich wie eine Auflistung. Aber ich würde sagen ein Punkt baut auf dem anderen auf. ==> RESOLVED
When within the FGR treatment the dissipative strokes last infinitely long, such that thermalization is perfect, 
the work extracted from one cycle~\eqref{EQ:worksys} and the heat absorbed from the cold reservoir~\eqref{EQ:heatsys}, respectively, 
can be computed as (see also App.~\ref{APP:2l_infslow})
\begin{align}\label{EQ:workheatFGR2level}
\Delta W_{\rm FGR}^\infty &= \frac{\omega_h-\omega_c}{2} \left[\tanh\left(\frac{\beta_c\omega_c}{2}\right)-\tanh\left(\frac{\beta_h\omega_h}{2}\right)\right]\,,\nn
\Delta Q_{\rm c,FGR}^\infty &= \frac{\omega_c}{2} \left[\tanh\left(\frac{\beta_h\omega_h}{2}\right)-\tanh\left(\frac{\beta_c\omega_c}{2}\right)\right]\,.
\end{align}
This limit is useful to find conditions under which work extraction ($\Delta W>0$) or cooling functionality ($\Delta Q_c>0$) is possible in principle and even optimal.

\begin{figure}
\includegraphics[width=0.45\textwidth]{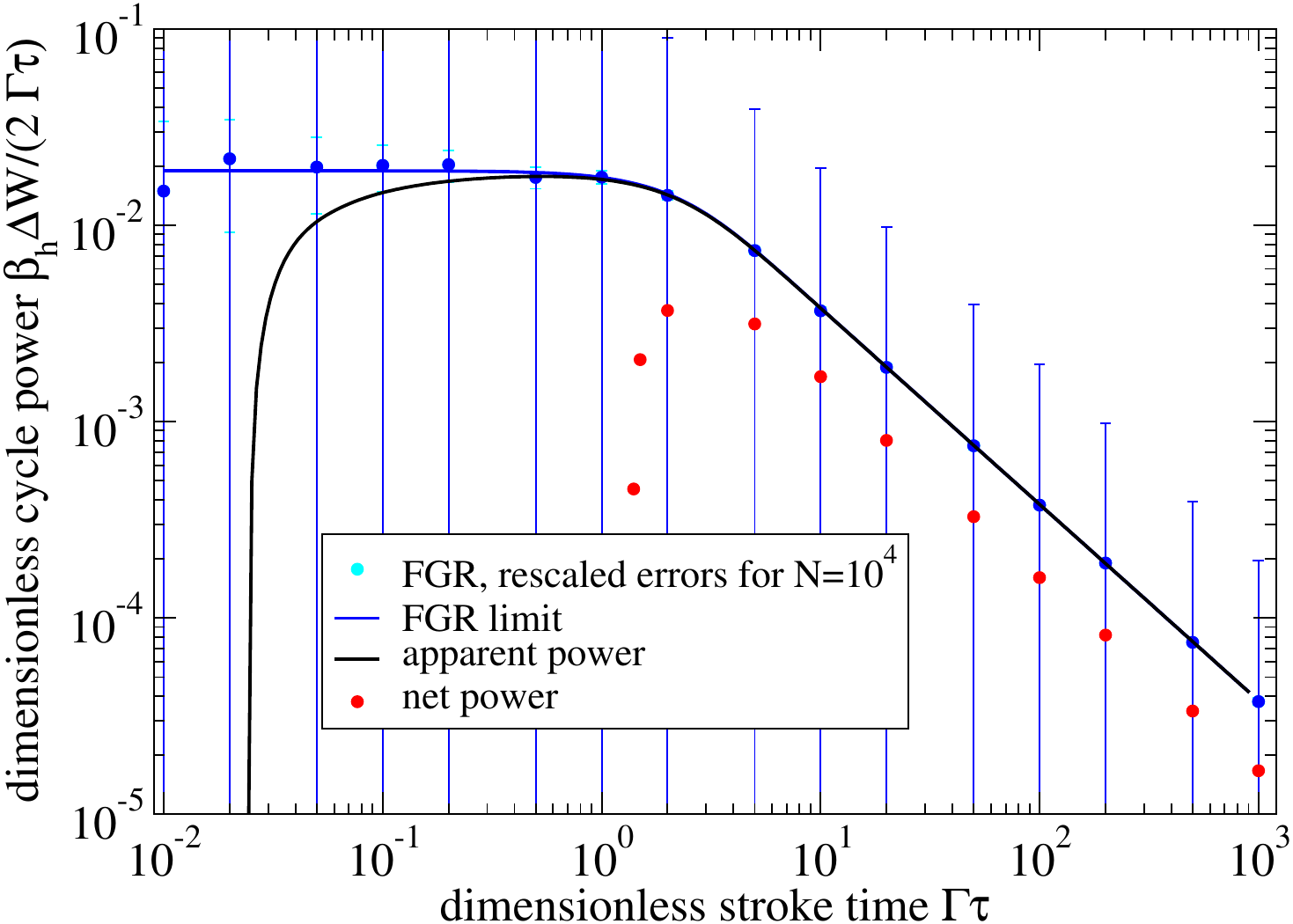}\\
\includegraphics[width=0.45\textwidth]{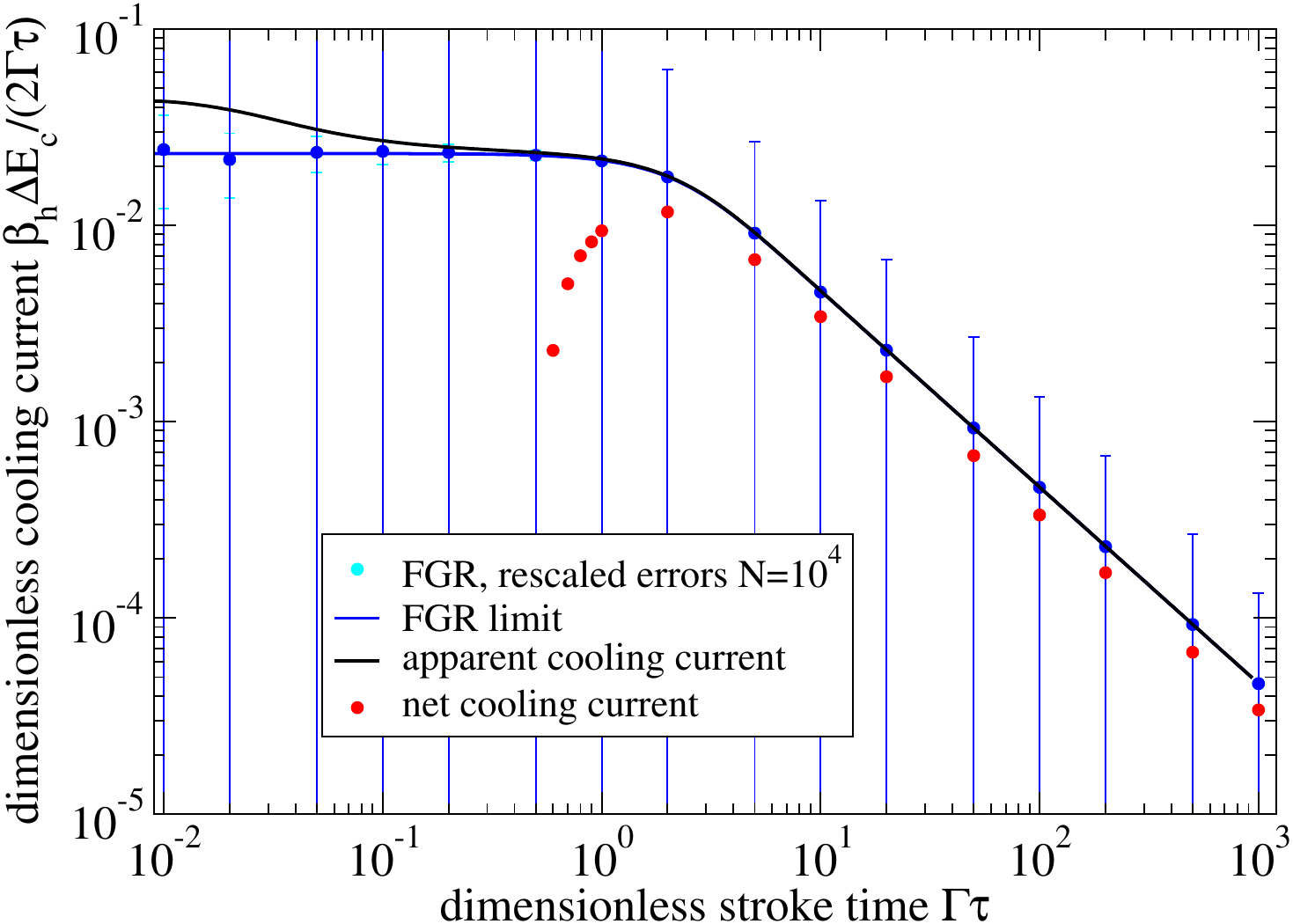}
\caption{\label{FIG:powercooling2level}
Engine performance per cycle time versus different stroke durations.
Blue curves result from a naive Fermi-Golden-Rule treatment also for finite-time cycles.
Black curves include the finite stroke durations but only consider the energetic balances of the system and neglect the contribution from control work.
Red symbols include the cost of control.
{\bf Top:} Extracted power for $\beta_h \omega_c = 1.05612$ and $\beta_h \omega_h = 1.86384$.
{\bf Bottom:} Cooling current for $\beta_h \omega_c = 0.426156$ and $\beta_h \omega_h = 10$.
Other parameters (for both plots) $\beta_c = 3 \beta_h$ and $\beta_h \Gamma = 0.01$.
For the FGR curves we also show mean and standard deviation obtained from $10^5$ trajectories (blue symbols).
}
\end{figure}

Still within an FGR treatment, equilibration with the reservoir will not be perfect for finite cycle-times, and a smaller limit cycle will emerge.
One may analytically solve the dynamical equations \new{for the limit cycle and finite stroke durations} (see App.~\ref{APP:2l_ftcycle}), and we then find with FGR rates that the extracted work~\eqref{EQ:worksys} and cooling heat~\eqref{EQ:heatsys} are simply reduced by a common factor (see App.~\ref{APP:2l_ftfgr} \new{and Ref.~\cite{dann2020a} for a generalization beyond frictionless working fluids})
\begin{align}\label{EQ:workheatFGR2levelred}
\Delta W_{\rm FGR}^\tau &= \Delta W_{\rm FGR}^\infty \tanh\left(\frac{\Gamma\tau}{2}\right)\,,\nn
\Delta Q_{\rm c,FGR}^\tau &= \Delta Q_{\rm c,FGR}^\infty \tanh\left(\frac{\Gamma\tau}{2}\right)\,.
\end{align}
This would formally imply that the extracted power $\Delta W_{\rm FGR}^\tau/(2\tau)$ and cooling current $\Delta Q_{\rm c,FGR}^\tau/(2\tau)$ would become maximal for very fast cycles $\tau\to 0$, compare the blue curves in Fig.~\ref{FIG:powercooling2level}.
There, we plot the power (top panel) and the cooling current (lower panel) versus the stroke duration for parameters that are optimal for the respective purpose in the infinitely slow limit, i.e., that
maximize~\eqref{EQ:workheatFGR2level} for the fixed temperature ratio of $\beta_c = 3 \beta_h$. 
Trajectory simulations (blue symbols with error bars) confirm this picture but also show that the performance of single systems is quite noisy.

One may be tempted to assess the engine performance using the traditional definitions of work~\eqref{EQ:worksys} and heat~\eqref{EQ:heatsys} but with the coarse-grained rates~\eqref{EQ:ratecg} instead, compare the black curves in Fig.~\ref{FIG:powercooling2level} and also App.~\ref{APP:2l_ftcg}.
This treatment would correctly reproduce the system dynamics (at least in the weak-coupling regime), but neglects the control work spent for the cycle operation. 
For cooling, this leads to the artifact that the apparent cooling current is above the FGR results (Fig.~\ref{FIG:powercooling2level} lower panel) at short strokes.
This increase however is just control work spent to increase the energy of the system and therefore does not contribute to cooling of the cold reservoir.

By considering coarse-grained rates~\eqref{EQ:ratecg} for the system and the associated average energetic changes in the reservoir~\eqref{EQ:reschangecg} one can address the net extracted work~\eqref{EQ:network} and the net cooling heat~\eqref{EQ:netheat}, compare the red symbols in Fig.~\ref{FIG:powercooling2level}.
The drastic difference between black curves and red symbols shows that the system energy changes are not sufficient to judge cycle efficiency: 
Not only do very fast cycles become dysfunctional, but also the performance of very slow cycles is reduced in comparison to the FGR limit as successive energy-preserving transitions (e.g. excitation-de-excitation processes) will lead to reservoir heating.
In the plots, we see that for the net (cooling) power there is an optimal stroke duration of $\Gamma\tau\approx 2$, which leads to a significant reduction of the maximum extractable power and cooling current in comparison to the FGR treatment.

%%%%%%%%%%%%%%%%%%%%%%%%%%%%%%%%%%%%%%%%%%%%%%%%%%%%%%%%%%%%%%%%%%%%%%%%%%%%%%%%
%%%%%%%%%%%%%%%%%%%%%%%%%%%%%%%%%%%%%%%%%%%%%%%%%%%%%%%%%%%%%%%%%%%%%%%%%%%%%%%%

%%%%%%%%%%%%%%%%%%%%%%%%%%%%%%%%%%%%%%%%%%%%%%%%%%%%%%%%%%%%%%%%%%%%%%%%%%%%%%%%
%%%%%%%%%%%%%%%%%%%%%%%%%%%%%%%%%%%%%%%%%%%%%%%%%%%%%%%%%%%%%%%%%%%%%%%%%%%%%%%
\section{Anisotropic 2d Ising model}\label{SEC:a2dim}
%%%%%%%%%%%%%%%%%%%%%%%%%%%%%%%%%%%%%%%%%%%%%%%%%%%%%%%%%%%%%%%%%%%%%%%%%%%%%%%%
%%%%%%%%%%%%%%%%%%%%%%%%%%%%%%%%%%%%%%%%%%%%%%%%%%%%%%%%%%%%%%%%%%%%%%%%%%%%%%%%

\begin{figure}
\includegraphics[width=0.4\textwidth]{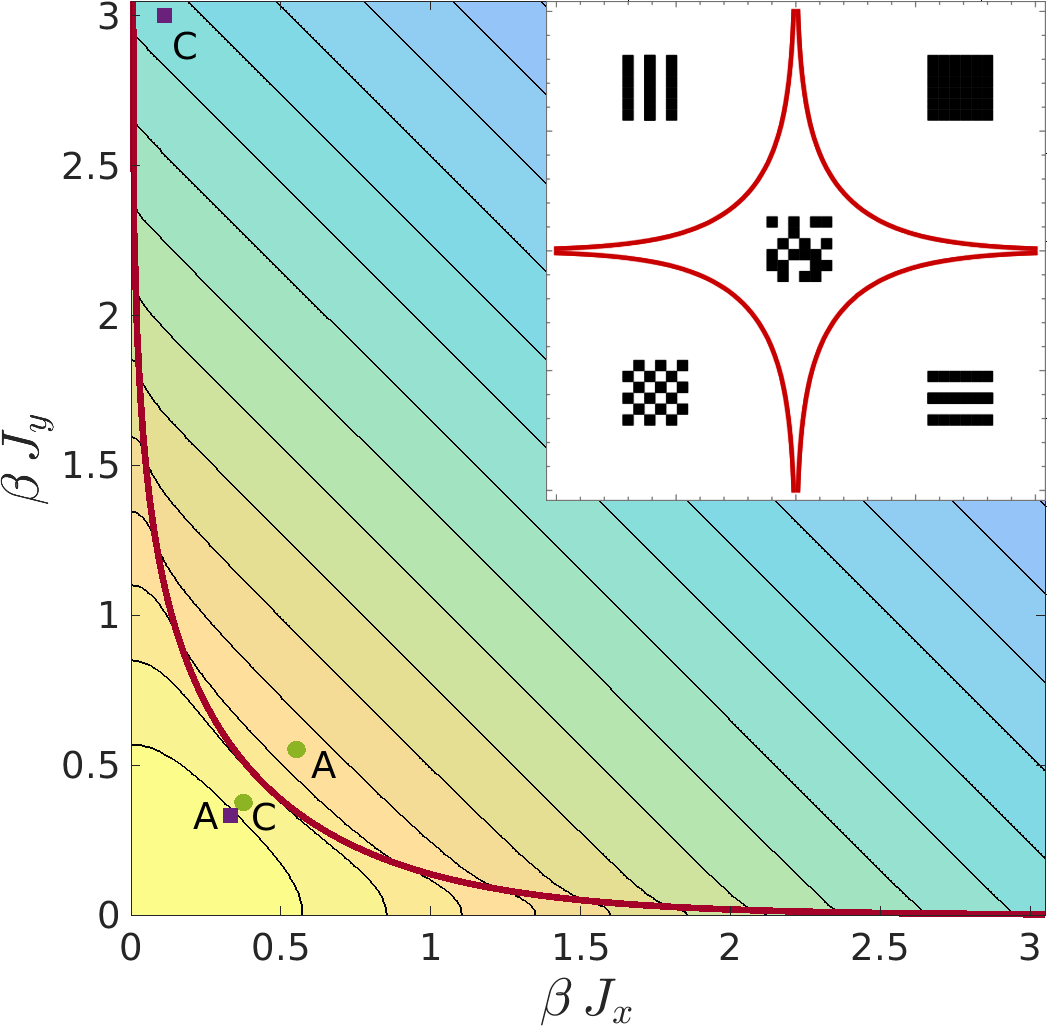}
\caption{\label{FIG:isingphasediagram}
Contour plot of the equilibrium energy per spin of the Ising model~\eqref{EQ:ising} \new{in the first quadrant $J_{x/y} \ge 0$} with phase transition~\eqref{EQ:phasetransition} marked red.
The dots mark the points of the Onsager cycle in equilibrium with the cold reservoir (A) or with the hot reservoir (C), where parameters $J_x^\nu$ and $J_y^\nu$ have been optimized for a fixed ratio of $\beta_c = 3 \beta_h$ to maximize work extraction (green circles) or to maximize cooling of the cold reservoir (purple squares). The purple $C$-square can be significantly shifted vertically without changing the cooling performance. For optimal performance, different phases of the working fluid need to be explored.
\new{The inset sketches an exemplary equilibrium configuration for all quadrants.}
}
\end{figure}

The anisotropic 2d classical Ising model
\begin{align}\label{EQ:ising}
H_S(t) = -J_x(t) \sum_{ij} \sigma^z_{ij} \sigma^z_{i+1,j} - J_y(t) \sum_{ij} \sigma^z_{ij} \sigma^z_{i,j+1}
\end{align}
has an exact equilibrium solution due to Onsager~\cite{onsager1944a}.
It predicts an order-disorder phase transition at
\begin{align}\label{EQ:phasetransition}
\sinh(2\beta \abs{J_x}) \sinh(2\beta \abs{J_y}) = 1\,,
\end{align}
with the disordered phase at the center $\beta J_{x/y} \approx 0$ (corresponding e.g. to random spin orientations at high temperatures) and ordered phases in the quadrants (homogeneous domains, domains with vertical stripes, checkerboard domains, and horizontal stripe domains, respectively, \new{as indicated in the inset of Fig.~\ref{FIG:isingphasediagram}}).
At the phase transition, the correlation lengths diverge with the Ising critical exponent $\nu=1$, but will for the anisotropic model in general depend on direction~\cite{mccoy1973}.
It is known~\cite{kogut1979a} that its partition function and hence its equilibrium properties can be mapped to the 1d quantum Ising model in a transverse field~\cite{sachdev2011}, and the investigations on the latter as a quantum working fluid~\cite{piccitto2022a,arezzo2024a} motivate us to consider the properties of this classical model as well.
The model can for example be implemented with silicon, where the spins map to the tilt of silicon dimers on the surface~\cite{brand2023a}.
Since energy-wise, the quadrants are equivalent, we concentrate on the first quadrant here and only consider $J_x, J_y>0$ \new{as in the main plot of} Fig.~\ref{FIG:isingphasediagram}.

\begin{figure}
%\begin{tabular}{cc}
\includegraphics[width=0.45\textwidth]{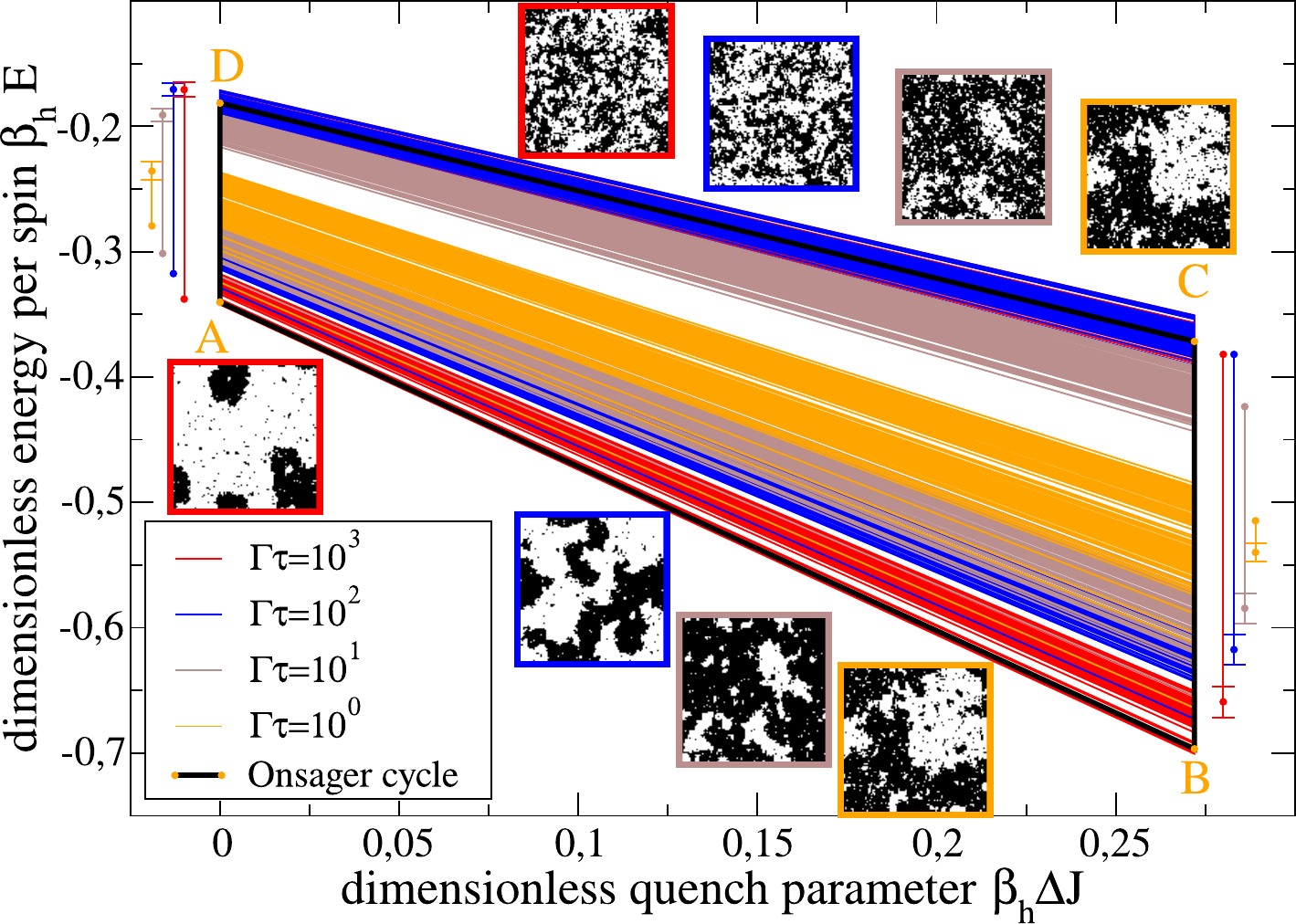}
\includegraphics[width=0.45\textwidth]{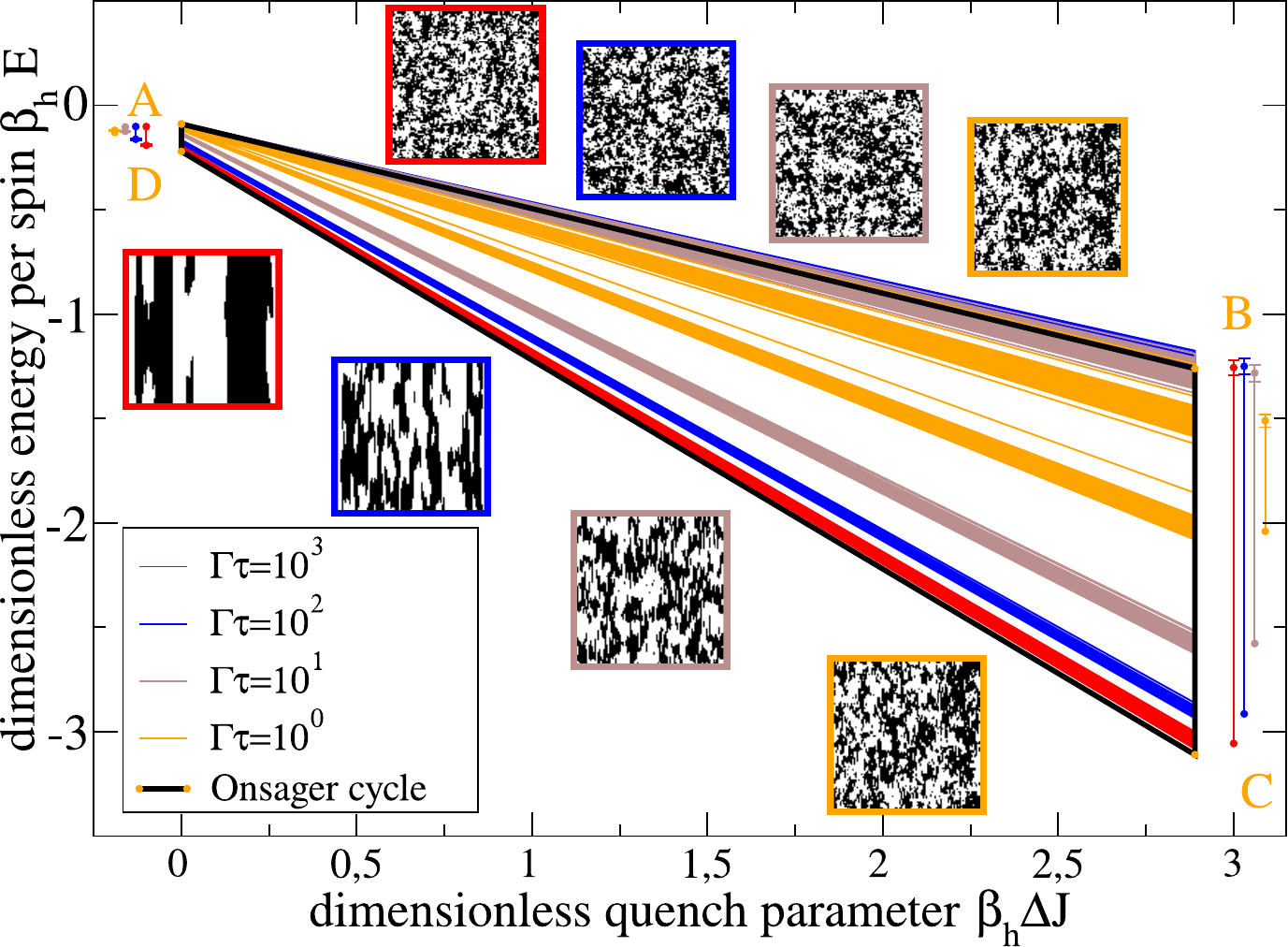}
%\end{tabular}
\caption{\label{FIG:cycleplot}
Onsager cycles (black with orange dots) and finite-time trajectory realizations for 100 cycles of a $100\times 100$ Ising spin working fluid and varying dissipative stroke lengths $\tau$ (red, blue, brown, orange), an initial equilibration time with the cold reservoir $\Gamma_c \tau_{\rm ini} = 10^3$ in the weak coupling regime  $\Gamma_\nu \beta_h = 0.01$.
The length of the side-bars denote the corresponding reservoir energy change in the reservoir, with error bars on the top indicating an increase and error bars at the bottom a reduction.
Snapshots framed with the trajectory color show the working fluid state in the last cycle during $A\to B$ or $C\to D$, respectively.
{\bf Top:} Work extraction cycle for optimized parameters in the Onsager cycle of $\beta_h J_x^c = \beta_h J_y^c = 0.1837$ and $\beta_h J_x^h = \beta_h J_y^h = 0.3760$.
For work extraction, the energy uptake from the hot reservoir must exceed the energy loss to the cold reservoir (\new{black,} red, blue, brown).
In the weak-coupling regime and for sufficiently large stroke times, the corresponding reservoir energy change is roughly minus the energy change of the system (i.e., control work is negligible). The fastest cycle (orange) is dysfunctional: On average, more heat is dissipated into the cold reservoir than is obtained from the hot one. 
{\bf Bottom:} Cooling for optimized parameters in the Onsager cycle of $\beta_h J_x^c = \beta_h J_y^c = 0.1105 = \beta_h J_x^h$ and $\beta_h J_y^h = 3$.
The cold reservoir is cooled when heat is taken out of the cold reservoir during $D\to A$. Only the slow cycles are operational (\new{black,} red, blue, brown), whereas the fastest is not (orange) as the cold reservoir is actually heated.
}
\end{figure}

First, for infinite stroke durations -- we use the term {\em Onsager cycle} to denote this limit -- we can obtain all energies by computing suitable derivatives of the 
partition function as we discuss in App.~\ref{APP:onsager}.
Numerical optimization then yields the parameters $J_{x/y}^\nu$ that are optimal for work extraction or cooling in one Onsager cycle.
As in the Onsager cycle, points $A$ and $C$ correspond to exact equilibrium states, we display these points in Fig.~\ref{FIG:isingphasediagram}, where we see that for optimal performance, the working medium has to explore different phases.
We also mention that the optimum for work extraction is pretty narrow, but the optimal value for cooling (purple $C$-square in Fig.~\ref{FIG:isingphasediagram}) can be moved vertically without substantial changes to the Onsager cycle performance (its optimal $J_y$-value is presumably at infinity).
The optimal Onsager cycles are depicted by the orange dots and black lines in Fig.~\ref{FIG:cycleplot}, where one can see that the orientation of the cycle for cooling is opposite to the orientation for work extraction.

Second, for finite-time cycles, none of the cycle points correspond to exact equilibrium states.
Nevertheless, by computing trajectories (see App.~\ref{APP:stochprop}) we can also obtain solutions of rate equations.
In contrast to the two level system with energy differences $\pm \omega_\nu$, already single spinflips allow for nine different system energy changes per jump
$\Omega\in\{\pm (4J_x^\nu +4 J_y^\nu), \pm (4 J_x^\nu - 4 J_y^\nu), \pm4 J_x^\nu, \pm4 J_y^\nu, 0\}$ that enter the coarse-grained rates~\eqref{EQ:ratecg}.
For each system energy change $\Omega$ we can also compute the associated average reservoir energy change~\eqref{EQ:reschangecg}.
To model the relaxation technically, we determine the associated waiting times and perform jumps such that on average, 
the solution of the rate equation is reproduced, see App~\ref{APP:stochprop}.
After an initial rather long equilibration with the cold reservoir, the Otto cycle protocol is repeated for 100 cycles.
In Fig.~\ref{FIG:cycleplot} we see that the trajectories (red, blue, brown, and orange) converge into a limit cycle that is for fast cycles significantly smaller than the Onsager cycle.
Furthermore, one also sees that convergence to the limit cycle requires more cycles for short stroke durations (orange) than for long stroke durations (red).
For slow cycles (red), the working fluid state is close to the expected equilibrium state (red framed snapshots) at points A and C, respectively, whereas for very fast cycles (orange and correspondingly framed snapshots) the differences between $A\to B$ and $C\to D$ are hardly visible.
To assess the actual performance of the cycle one has to consider the true heat taken from the reservoirs~\eqref{EQ:reschangecg}.
Its average value over the 100 cycles is indicated by the side bars in Fig.~\ref{FIG:cycleplot} with their length denoting the energetic change and where error bars at the top indicate an energy increase and error bars at the bottom an energy decrease of the corresponding reservoir.
From that, we see e.g. from the lengths of the orange sidebars in Fig.~\ref{FIG:cycleplot} top panel that more heat is dissipated in the cold reservoir than is taken from the hot one.
Thus, to run the cycle that fast, the necessary control work overturns the work extracted (which we get from the system perspective only).
An analogous failure is observed for the fast cycle aimed at cooling (orange curve in the bottom panel), where the invested control work on average leads to slight heating of the cold reservoir.

\begin{figure}
%\begin{tabular}{cc}
\includegraphics[width=0.45\textwidth]{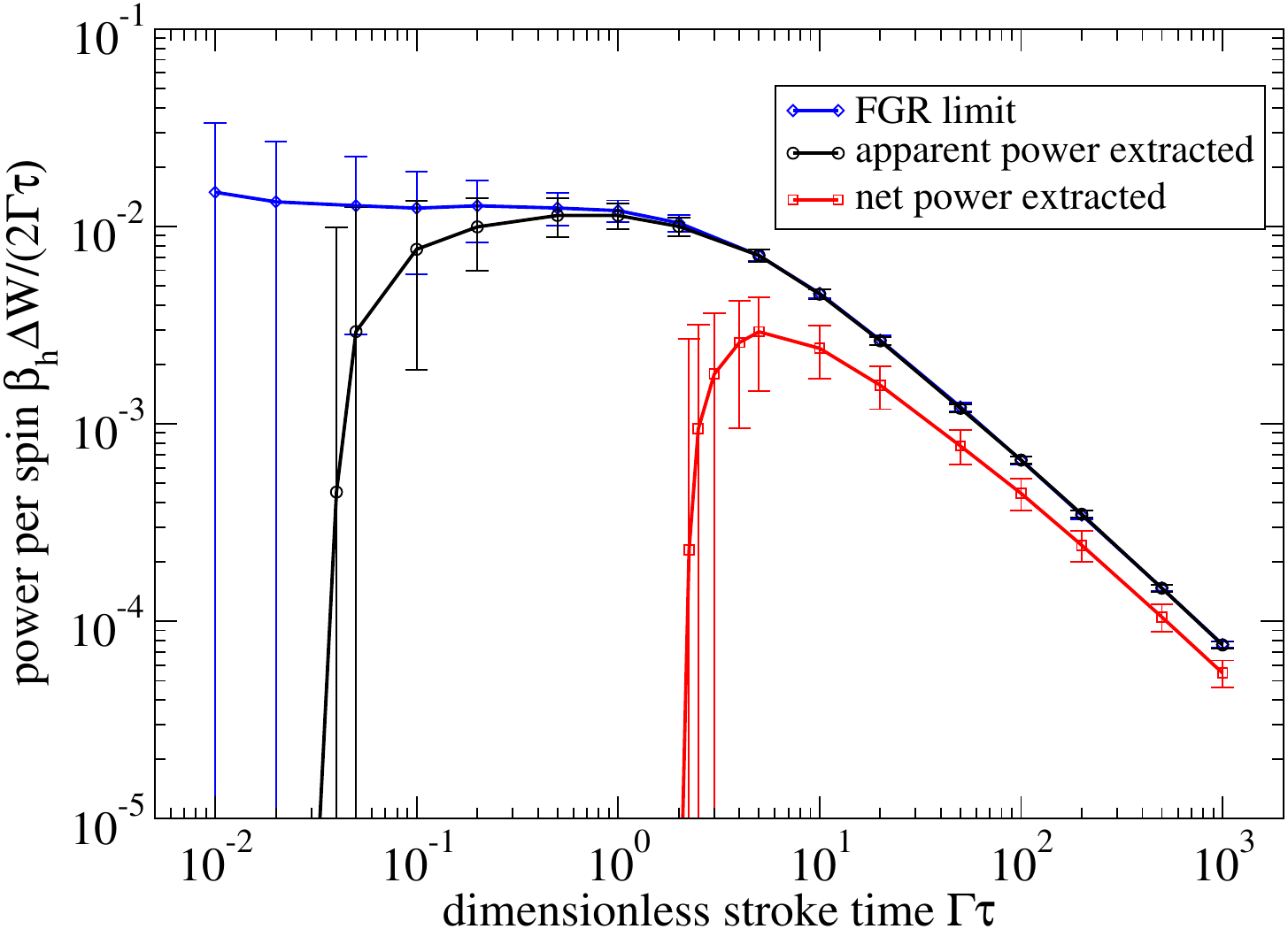}
\includegraphics[width=0.45\textwidth]{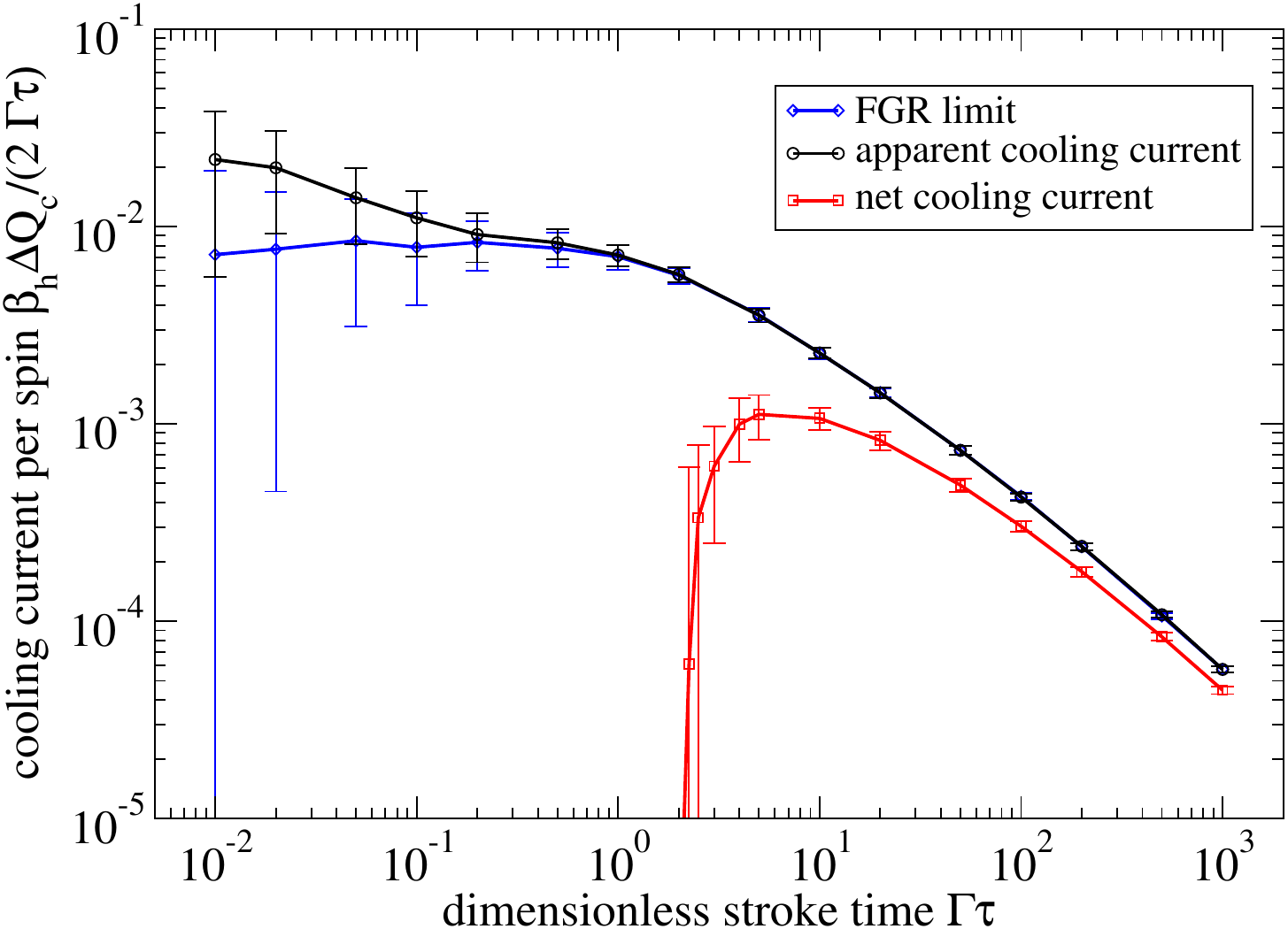}
%\end{tabular}
\caption{\label{FIG:performanceplot}
Engine performance per cycle time versus different stroke durations in the limit cycle regime (error bars denote the statistical error obtained from 100 cycles after an initial relaxation protocol of $N_{\rm cyc}^{\rm ini}$ cycles with $N_{\rm cyc}^{\rm ini} \Gamma \tau \ge 100$).
Blue curves just use Fermi golden rule rates independent of the cycle time, black curves use stroke-time-dependent rates but neglect the control work spent, whereas red curves use coarse-grained rates and include the control work for coupling and decoupling. A realistic assessment (red) leads to a significant loss in cycle performance.
{\bf Top:} Extracted power for $\beta_h J_x^c = \beta_h J_y^c = 0.1837$ and $\beta_h J_x^h = \beta_h J_y^h = 0.3760$.
{\bf Bottom:} Cooling current for $\beta_h J_x^c = \beta_h J_y^c = 0.1105 = \beta_h J_x^h$ and $\beta_h J_y^h = 3$.
Other parameters (identical for both plots): $\beta_c = 3 \beta_h$,  $\beta_h \Gamma_\nu = 0.01$, $\beta_h \delta_\nu = 10^3$.
}
\end{figure}

These findings motivate us to include the control work~\eqref{EQ:controlwork} of coupling and decoupling into the discussion and
to study the cycle performance as a function of stroke duration. 
In Fig.~\ref{FIG:performanceplot} we contrast the FGR treatment (blue, based on FGR rates~\eqref{EQ:rates_fgr} with standard work~\eqref{EQ:worksys} and heat~\eqref{EQ:heatsys} definitions), the apparent coarse-grained treatment (black, using coarse-grained rates~\eqref{EQ:ratecg} but still standard work~\eqref{EQ:worksys} and heat~\eqref{EQ:heatsys} definitions), 
and the net coarse-grained treatment (red, using rates for system~\eqref{EQ:ratecg} and reservoir changes~\eqref{EQ:reschangecg} with net work~\eqref{EQ:network} and net heat~\eqref{EQ:netheat} definitions), for both the power extraction (top panel) and cooling current (bottom panel).
Qualitatively, the results are quite analogous to those of Fig.~\ref{FIG:powercooling2level}, and the involved energies per spin as well as the timescales are comparable.
The relevant net work and net cooling current both show a pronounced maximum at $\Gamma \tau \approx 5$, for too slow cycles the net performance goes to zero, and too fast cycles are simply dysfunctional.
We also mention that the discrepancy between red and other curves would become larger for larger system-reservoir coupling strengths.
For example, already for $\Gamma_\nu \beta_\nu = 0.1$, none of the cycles would be functional when the net work/heat is considered (not shown).
Furthermore, we also see that the error bars of the Ising model are significantly smaller than the error bars of the two level system.
This however is just due to the significantly larger system: If we considered $10^4$ independently operating two-level systems, we would have to rescale the error bars from Fig.~\ref{FIG:powercooling2level} by $10^{-2}$, and the errors would become comparable.
Finally, we think that the increase of error bars in regions of reduced currents for fast strokes is just due to bistable behaviour~\cite{jordan2004a}:
Some cycles will yield a positive output, others a negative one, which naturally goes along with increased variance.

%%%%%%%%%%%%%%%%%%%%%%%%%%%%%%%%%%%%%%%%%%%%%%%%%%%%%%%%%%%%%%%%%%%%%%%%%%%%%%%%
%%%%%%%%%%%%%%%%%%%%%%%%%%%%%%%%%%%%%%%%%%%%%%%%%%%%%%%%%%%%%%%%%%%%%%%%%%%%%%%%
\section{Conclusions} 
%%%%%%%%%%%%%%%%%%%%%%%%%%%%%%%%%%%%%%%%%%%%%%%%%%%%%%%%%%%%%%%%%%%%%%%%%%%%%%%%
%%%%%%%%%%%%%%%%%%%%%%%%%%%%%%%%%%%%%%%%%%%%%%%%%%%%%%%%%%%%%%%%%%%%%%%%%%%%%%%%

The main purpose of our study was to investigate the use of a 2d Ising model as a quantum working fluid in a finite-stroke heat engine. 
Although the working fluid explores different phases for optimal performance, a frictionless drive does not induce critical slow-down, and the 
resulting cyclic operation shows characteristics that are similar to an uncritical two-level system.

Additionally, we pointed out the importance of assessing reservoir energetic changes in finite-time cyclic quantum heat engines.
For finite times, even in the weak-coupling regime \new{and for frictionless working fluids}, an analysis of quantum Otto cycles taking the control
work contributions into account reveals the existence of an optimal stroke duration time.
When cycles are driven much faster, the control work spent for coupling and decoupling to reservoirs exceeds the gain.
Also for slower cycles, (net) zero energy transitions may lead to additional reservoir heating.

We observed these effects for finite-time cycles already in the weak-coupling regime.
In the regimes of stronger system-reservoir couplings, the control work can be expected to become even more relevant~\cite{newman2017a,newman2020a}.
Exactly solvable models~\cite{pozas_kerstjens2018a} may prove useful to calibrate the theoretical methods in this regime, 
which is an interesting field of further research.

%%%%%%%%%%%%%%%%%%%%%%%%%%%%%%%%%%%%%%%%%%%%%%%%%%%%%%%%%%%%%%%%%%%%%%%%%%%%%%%%
%%%%%%%%%%%%%%%%%%%%%%%%%%%%%%%%%%%%%%%%%%%%%%%%%%%%%%%%%%%%%%%%%%%%%%%%%%%%%%%%
\section{Acknowledgments} 
%%%%%%%%%%%%%%%%%%%%%%%%%%%%%%%%%%%%%%%%%%%%%%%%%%%%%%%%%%%%%%%%%%%%%%%%%%%%%%%%
%%%%%%%%%%%%%%%%%%%%%%%%%%%%%%%%%%%%%%%%%%%%%%%%%%%%%%%%%%%%%%%%%%%%%%%%%%%%%%%%

The authors thank C. Henkel, A. Hucht, K. Hovhannisyan, and P. Kratzer for discussions and the DFG
(Project ID 278162697 -- SFB 1242) for financial support~\footnote{The data that support the findings of this article are openly available [\url{https://doi.org/10.14278/rodare.3875}].}.

%%%%%%%%%%%%%%%%%%%%%%%%%%%%%%%%%%%%%%%%%%%%%%%%%%%%%%%%%%%%%%%%%%%%%%%%%%%%%%%%
%%%%%%%%%%%%%%%%%%%%%%%%%%%%%%%%%%%%%%%%%%%%%%%%%%%%%%%%%%%%%%%%%%%%%%%%%%%%%%%%
\bibliographystyle{unsrt}
\bibliography{references.bib}
%%%%%%%%%%%%%%%%%%%%%%%%%%%%%%%%%%%%%%%%%%%%%%%%%%%%%%%%%%%%%%%%%%%%%%%%%%%%%%%%
%%%%%%%%%%%%%%%%%%%%%%%%%%%%%%%%%%%%%%%%%%%%%%%%%%%%%%%%%%%%%%%%%%%%%%%%%%%%%%%%

\newpage
\appendix

%%%%%%%%%%%%%%%%%%%%%%%%%%%%%%%%%%%%%%%%%%%%%%%%%%%%%%%%%%%%%%%%%%%%%%%%%%%%%%%%
%%%%%%%%%%%%%%%%%%%%%%%%%%%%%%%%%%%%%%%%%%%%%%%%%%%%%%%%%%%%%%%%%%%%%%%%%%%%%%%%
\section{Two-level system}\label{APP:two-level}
%%%%%%%%%%%%%%%%%%%%%%%%%%%%%%%%%%%%%%%%%%%%%%%%%%%%%%%%%%%%%%%%%%%%%%%%%%%%%%%%
%%%%%%%%%%%%%%%%%%%%%%%%%%%%%%%%%%%%%%%%%%%%%%%%%%%%%%%%%%%%%%%%%%%%%%%%%%%%%%%%

\subsection{Infinitely slow cycle}\label{APP:2l_infslow}

For infinitely long equilibration times, the states at points $A$ and $C$ of the cycle will just be thermal ones.
The partition function $Z(\beta) = 2 \cosh(\beta\omega/2)$ yields the energy $E_{\rm th} = -\frac{\omega}{2} \tanh\left(\frac{\beta\omega}{2}\right)$ in a thermal state at inverse temperature $\beta$.
We can calculate all energies along an idealized cycle
\begin{align}
E_A &= -\frac{\omega_c}{2} \tanh\left(\frac{\beta_c\omega_c}{2}\right)\,,\nn
E_B &= -\frac{\omega_h}{2} \tanh\left(\frac{\beta_c\omega_c}{2}\right)\,,\nn
E_C &= -\frac{\omega_h}{2} \tanh\left(\frac{\beta_h\omega_h}{2}\right)\,,\nn
E_D &= -\frac{\omega_c}{2} \tanh\left(\frac{\beta_h\omega_h}{2}\right)\,,
\end{align}
where $E_A$ and $E_C$ just correspond to the thermal expectation values, and where we have due to the simple structure of the Hamiltonian deduced the other energies from
$E_B = \frac{\omega_h}{\omega_c} E_A$ and $E_D = \frac{\omega_c}{\omega_h} E_C$.
From this, we in turn obtain the extracted work and the heat entering from the cold reservoir as in Eq.~\eqref{EQ:workheatFGR2level} in the main text.
The heat entering from the hot reservoir becomes
\begin{align}
\Delta Q_{\rm h,FGR} &= \frac{\omega_h}{2} \left[\tanh\left(\frac{\beta_c\omega_c}{2}\right)-\tanh\left(\frac{\beta_h\omega_h}{2}\right)\right]\,.
\end{align}

For the extraction of work, it follows that 
the efficiency is $\eta=\left(1-\frac{\omega_c}{\omega_h}\right)\Theta(\Delta W) \le \eta_{\rm Carnot}$, where the last inequality follows by case distinction: When $\omega_h > \omega_c$, we have to fulfil $\beta_c \omega_c > \beta_h \omega_h$, such that the extraction of work is enabled when $\omega_h \in [\omega_c, \omega_c \frac{\beta_c}{\beta_h}]$.
To the contrary, when $\omega_h < \omega_c$, we would also reqire $\beta_c \omega_c < \beta_h \omega_h$, which can for negative $\omega_\nu$ be achieved as $\omega_h \in [\omega_c \frac{\beta_c}{\beta_h}, \omega_c]$.

To cool the cold reservoir, we have to invest work, leading to a coefficient of performance $\kappa = \Delta Q_c/(-\Delta W) \Theta(\Delta Q_c) = \frac{1}{\frac{\omega_h}{\omega_c}-1} \Theta(\Delta Q_c)$.
The additional constraint that heat should be extracted from the cold reservoir leads for $\omega_c>0$ to cooling in the window $\omega_h \in [\beta_c \omega_c/\beta_h, \infty)$, or for $\omega_c<0$ to cooling in the window $\omega_h \in (-\infty, \beta_c \omega_c/\beta_h]$.

Thus, we always have $\omega_h \omega_c > 0$ for cooling or work extraction, and level crossings (with $\omega_h \omega_c < 0$) would not allow the extraction of work or cooling of the cold reservoir with the cycle operation.

\subsection{Finite-time cycles}\label{APP:2l_ftcycle}

When we do not quench through a crossing during $A\to B$ or $C\to D$ (as necessary for useful function), a system starting in an excited state will remain so during the isentropic quenches.
For the finite-time dissipative strokes, the two probabilities $P_+$ for being in the excited state and $P_-=1-P_+$ for being in the ground state will thus obey 
(excitation rate $R_\uparrow$ and deexcitation rate $R_\downarrow$ can in principle still depend on the stroke duration)
\begin{align}
\dot P_{+} &= R_\uparrow P_{-} - R_\downarrow P_{+}\,,\qquad
\dot P_{-} = -\dot P_{+}\,,
\end{align}
which can be directly solved
\begin{align*}
P_{+}(t) &= e^{-(R_\uparrow+R_\downarrow)t} P_+^0 + \frac{R_\uparrow}{R_\downarrow+R_\uparrow} \left(1 - e^{-(R_\uparrow+R_\downarrow)t}\right)\,.
\end{align*}
This allows to construct the full time-dependent solution by concatenating the above evolution along the dissipative strokes with constant populations along the unitary strokes and for alternating thermal reservoirs.
For example, when $P_{+,n}^A(t)$ denotes the probability to be in the excited state ($+$) during the stroke $D\to A$ in the $n$-th repetition of the cycle, it follows from the constant populations during the unitary strokes that the initial occupation of each (dissipative) stroke is given by the final occupation of the previous (dissipative) stroke
\begin{align}
P_{+,n}^C(0) = P_{+,n}^A(\tau_c)\,,\qquad
P_{+,n+1}^A(0) = P_{+,n}^C(\tau_h)\,.
\end{align}
When the system reaches a limit cycle, we may drop the index $n$, and the above two equations constitute a linear system that can be solved for the limit cycle probabilities (overbars)
$\bar P_+^A \equiv \lim\limits_{n\to\infty} P_{+,n}^A(\tau_c)$ and 
$\bar P_+^C \equiv \lim\limits_{n\to\infty} P_{+,n}^C(\tau_h)$.
For example, setting the duration of the dissipative strokes equal $\tau_c=\tau_h=\tau$, the limit cycle probabilities are obtained by solving
\begin{align}\label{EQ:limcycprobs}
\bar P_{+}^A &= e^{-(R_\uparrow^c+R_\downarrow^c) \tau} \bar P_+^C + \frac{R_\uparrow^c}{R_\downarrow^c+R_\uparrow^c} \left(1 - e^{-(R_\uparrow^c+R_\downarrow^c)\tau}\right)\,,\nn
\bar P_{+}^C &= e^{-(R_\uparrow^h+R_\downarrow^h) \tau} \bar P_+^A + \frac{R_\uparrow^h}{R_\downarrow^h+R_\uparrow^h} \left(1 - e^{-(R_\uparrow^h+R_\downarrow^h)\tau}\right)\,,
\end{align}
where $R^\nu_\uparrow$ and $R^\nu_\downarrow$ are the excitation and decay rates when coupled to reservoir $\nu$, respectively.
From these, we can then compute the system energies at all points of the limit cycle via
\begin{align}
\bar E_A &= \frac{\omega_c}{2} [2 \bar P_{+}^A - 1]\,,\qquad
\bar E_B = \frac{\omega_h}{2} [2 \bar P_{+}^A - 1]\,,\nn
\bar E_C &= \frac{\omega_h}{2} [2 \bar P_{+}^C - 1]\,,\qquad
\bar E_D = \frac{\omega_c}{2} [2 \bar P_{+}^C - 1]\,,
\end{align}
and from that we find that the apparent work~\eqref{EQ:worksys} per limit cycle and the heat entering the system~\eqref{EQ:heatsys} via the cold thermalization stroke, respectively, are limited by the difference of excited state populations via
\begin{align}
\Delta \bar W = (\bar P_+^C - \bar P_+^A)(\omega_h-\omega_c)\,,\qquad
\Delta \bar Q_c = (\bar P_+^A - \bar P_+^C) \omega_c\,.
\end{align}

\subsection{Fermi golden rule rates}\label{APP:2l_ftfgr}

In the specific case where the rates are independent of the cycle time given by $R^c_\uparrow = \Gamma f_c$, $R^c_\downarrow = \Gamma (1-f_c)$, $R^h_\uparrow = \Gamma f_h$, and $R^h_\downarrow = \Gamma (1-f_h)$ with 
$f_\nu = [e^{\beta_\nu \omega_\nu}+1]^{-1}$, we obtain that work and transferred heats are all reduced by the same factor
as in Eq.~\eqref{EQ:workheatFGR2levelred} in the main text
\begin{align}
0 \le \tanh\left(\frac{\Gamma \tau}{2}\right) \le 1\,,
\end{align}
such that efficiency and coefficient of performance would just remain the same.
This factor shows that for very large cycle times, the previous results are recovered, but in this limit the power (work by cycle time) and also the cooling current (heat per period by cycle time) will drop to zero.
Additionally, constant rates would suggest that to maximize power, it would be much more favorable to operate in the regime where $\Gamma \tau \ll 1$, where the power roughly approaches $P \approx \frac{\Gamma}{4} \Delta W$, and the cooling current approaches $J_c \approx \frac{\Gamma}{4} \Delta Q_c$.
For this limit, we can also calculate the second moment of the work
\begin{align}
\expval{\Delta W^2} &= \expval{(E_A-E_B+E_C-E_D)^2}\\
&= \left<\left[\left(1-\frac{\omega_h}{\omega_c}\right)E_A + \left(1-\frac{\omega_c}{\omega_h}\right) E_C\right]^2\right>\nn
&= (\omega_h-\omega_c)^2/2\nn
&\qquad + \left(1-\frac{\omega_h}{\omega_c}\right)\left(1-\frac{\omega_c}{\omega_h}\right)\expval{E_A E_C + E_C E_A}\,.\nonumber
\end{align}
Eventually, this also yields the second cumulant of the work
\begin{align}
\langle\langle \Delta W^2 \rangle\rangle &= \frac{2 \left(e^{\Gamma \tau}-1\right) (\omega_h-\omega_c)^2}{\left(1+e^{\Gamma \tau}\right)^2} [f_c+f_h - 2 f_c f_h]\nn
&\qquad+\frac{\left(1-e^{-\Gamma \tau}\right)\left(1+e^{+2\Gamma \tau}\right) (\omega_h-\omega_c)^2}{\left(1+e^{\Gamma \tau}\right)^2}\times\nn
&\qquad\times [ f_c(1-f_c)+f_h(1-f_h)]\,,
\end{align}
which is always positive and for slow cycles $\Gamma \tau\to\infty$ approaches $\langle\langle \Delta W^2 \rangle\rangle \to (\omega_h-\omega_c)^2[f_c(1-f_c)+f_h(1-f_h)]$, which is just the result that one would get by treating the energies at $A$ and $C$ as statistically independent, such that their cumulants add up.
As a further sanity check, we also mention that when converting this into the error of the power, this also matches the blue error bars obtained from trajectories in Fig.~\ref{FIG:powercooling2level} top panel.

\subsection{Finite-time rates}\label{APP:2l_ftcg}

We can repeat the above analysis for realistic coarse-grained rates~\eqref{EQ:ratecg}, from which with exploiting symmetries of the Fermi function $f_\nu(\omega) = [e^{\beta_\nu\omega}+1]^{-1}$ we can write the excitation and deexcitation rates as
\begin{align}
R^{\nu,\tau}_\uparrow &= \int \Gamma_\nu(\omega) f_\nu(\omega) \frac{\tau}{2\pi} {\rm sinc}^2\left[\frac{(\omega-\omega_\nu) \tau}{2}\right] d\omega\,,\nn
R^{\nu,\tau}_\downarrow &= \int \Gamma_\nu(\omega) \frac{\tau}{2\pi} {\rm sinc}^2\left[\frac{(\omega-\omega_\nu) \tau}{2}\right] d\omega - R^{\nu,\tau}_\uparrow\,,
\end{align}
which explicitly depend on the stroke duration $\tau$.
Assuming for simplicity equal stroke durations, the wide-band limit $\Gamma_\nu(\omega)=\Gamma_\nu$, and also equal coupling strengths $\Gamma = R^{\nu,\tau}_\downarrow + R^{\nu,\tau}_\uparrow$ to the reservoirs, we obtain 
for the apparent work per cycle and apparent heat uptake from the cold reservoir
\begin{align}
\expval{\Delta W}^\tau_{\rm app} &= \frac{(R^{h,\tau}_\uparrow-R^{c,\tau}_\uparrow)(\omega_h-\omega_c)}{\Gamma} \tanh\left(\frac{\Gamma\tau}{2}\right)\,,\nn
\expval{\Delta Q_c}^\tau_{\rm app} &= \frac{(R^{c,\tau}_\uparrow - R^{h,\tau}_\uparrow)\omega_c}{\Gamma} \tanh\left(\frac{\Gamma\tau}{2}\right)\,,
\end{align}
which shows that for very short cycle times, both quantities scale quadratically in $\tau$, such that also the power and cooling current will tend to zero for very short stroke durations $\tau$ (such a turnover would become visible in Fig.~\ref{FIG:powercooling2level} lower panel for even shorter stroke times).

For finite cycle times $\tau$, the system energy changes do not correctly reflect the energetic changes of the reservoir.
Instead, when while coupled to reservoir $\nu$, the system undergoes the energy change $\Omega \in \{-\omega_\nu, +\omega_\nu\}$, the actual associated energy change~\eqref{EQ:reschangecg} of the reservoir would be
\begin{align}
\Delta E^{\nu,\tau}_\Omega = -\frac{\int \omega \Gamma_\nu(\omega) f_\nu(\omega) \frac{\tau}{2\pi} {\rm sinc}^2\left[\frac{(\omega-\Omega)\tau}{2}\right] d\omega}
{\int\Gamma_\nu(\omega) f_\nu(\omega) \frac{\tau}{2\pi} {\rm sinc}^2\left[\frac{(\omega-\Omega)\tau}{2}\right] d\omega}\,,
\end{align}
which for $\tau\to\infty$ just becomes $-\Omega$.
Thus, even a system undergoing no net transition at all (e.g. after one excitation and one de-excitation) will lead to heating of the reservoir.
To get the average net work extracted or average net cooling heat along the limit cycle, we have to sample over possible trajectories (compare App.~\ref{APP:stochprop}) with the coarse-grained rates and track the average reservoir energy changes accordingly, compare the red symbols Fig.~\ref{FIG:powercooling2level}.
The net result is that for $\tau\to 0$, fewer transitions per cycle happen and even for a low temperature reservoir excitation processes become just as likely as deexcitation processes:
No work can be extracted, no reservoir can be cooled, and all cycles are dysfunctional.
%

%%%%%%%%%%%%%%%%%%%%%%%%%%%%%%%%%%%%%%%%%%%%%%%%%%%%%%%%%%%%%%%%%%%%%%%%%%%%%%%%
%%%%%%%%%%%%%%%%%%%%%%%%%%%%%%%%%%%%%%%%%%%%%%%%%%%%%%%%%%%%%%%%%%%%%%%%%%%%%%%%
\section{Coarse-grained transition rates}\label{APP:coarsegraining}
%%%%%%%%%%%%%%%%%%%%%%%%%%%%%%%%%%%%%%%%%%%%%%%%%%%%%%%%%%%%%%%%%%%%%%%%%%%%%%%%
%%%%%%%%%%%%%%%%%%%%%%%%%%%%%%%%%%%%%%%%%%%%%%%%%%%%%%%%%%%%%%%%%%%%%%%%%%%%%%%%

\subsection{General derivation}\label{APP:general_derivation}

For finite stroke times $\tau$ and weak coupling strengths, time-dependent perturbation theory tells us that the system-bath density matrix can for an interaction $H_I$ be written as (bold symbols indicate the interaction picture)
\begin{align}
\f{\rho}(t+\tau) &= \f{\rho}(t) - \ii \left[\int\limits_t^{t+\tau} dt_1 \f{H_I}(t_1), \f{\rho}(t)\right]\\
&\qquad+ \int\limits_t^{t+\tau} dt_1 dt_2 \f{H_I}(t_1) \f{\rho}(t) \f{H_I}(t_2)\nn
&\qquad-\int\limits_t^{t+\tau} dt_1 dt_2 \Theta(t_1-t_2) \f{H_I}(t_1) \f{H_I}(t_2) \f{\rho}(t)\nn
&\qquad-\int\limits_t^{t+\tau} dt_1 dt_2 \Theta(t_2-t_1) \f{\rho}(t) \f{H_I}(t_1) \f{H_I}(t_2)\,.\nonumber
\end{align}
Thus, under the usual assumptions (specifically, assuming that at the beginning of the dissipative strokes $\f{\rho}(t) = \f{\rho_S}(t) \otimes \bar\rho_B$ with $[\bar\rho_B, H_B]=0$, writing furthermore the interaction Hamiltonian with hermitian system and bath coupling operators as $\f{H_I}(t) = \sum_\alpha \f{S_\alpha}(t) \otimes \f{B_\alpha}(t)$, the correlation function as $C_{\alpha\beta}(t_1-t_2)=\trace{\f{B_\alpha}(t_1) \f{B_\beta}(t_2) \bar\rho_B}$, and assuming that $\trace{\f{B_\alpha}(t) \bar\rho_B}=0$), the propagated system
density matrix becomes
\begin{align}
\f{\rho_S}(t+\tau) &= \f{\rho_S}(t)+ \int\limits_t^{t+\tau} dt_1 dt_2 \sum_{\alpha\beta} C_{\alpha\beta}(t_1-t_2)\times\nn
&\times \Big[
\f{S_\beta}(t_2) \f{\rho_S}(t) \f{S_\alpha}(t_1)\nn
&\qquad- \Theta(t_1-t_2) \f{S_\alpha}(t_1) \f{S_\beta}(t_2) \f{\rho_S}(t)\nn
&\qquad- \Theta(t_2-t_1) \f{\rho_S(t)} \f{S_\alpha}(t_1) \f{S_\beta}(t_2) \Big]\,.
\end{align}
In general, the associated coarse-grained dissipator ${\cal L}_\tau$ defined by ${\cal L}_\tau \f{\rho_S}(t) = \frac{\f{\rho_S}(t+\tau)-\f{\rho_S}(t)}{\tau}$ need not decouple the evolution of 
populations and coherences in the system energy eigenbasis.
Specific system-reservoir interactions may however lead to such a decoupled evolution.
For example, under the plausible assumption that each spin $(ij)$ of a 2d spin working fluid~\eqref{EQ:ising} hosts a single spinflip coupling operator $S_\alpha \to S_{ij}=\sigma^x_{ij}$ to its separate reservoir $C_{\alpha\beta}(\tau)\to
C_{k\ell,ij}(\tau) = \delta_{ki} \delta_{\ell j} C_{ij}(\tau)$ with correlation function as detailed below in App.~\ref{APP:specific_example}, it follows that for any two system energy eigenstates $\ket{a}$ and $\ket{b}$ and for an initially diagonal state 
$\f{\rho_S}(t) = \sum_c \rho_S^{cc}(t) \ket{c}\bra{c}$
one has for the time-evolved matrix elements $\rho_S^{ab}(t+\tau) \equiv \bra{a} \f{\rho_S}(t+\tau) \ket{b}$ the relations
\begin{align}
\rho_S^{ab}(t+\tau) &= \delta_{ab} \rho_S^{aa}(t) + \delta_{ab} \int\limits_t^{t+\tau} dt_1 dt_2 \sum_{ij} C_{ij}(t_1-t_2)\times\nn
&\qquad\times \Big[e^{-\ii (E_a-E_{F_{ij}^a})(t_1-t_2)} \rho_S^{F_{ij}^a F_{ij}^a}\nn
&\qquad\qquad- e^{+\ii (E_a-E_{F_{ij}^a})(t_1-t_2)} \rho_S^{aa}(t)\Big]\,,
\end{align}
where $\ket{F_{ij}^a} \equiv \sigma^x_{ij} \ket{a}$.
In other words, in this case initially diagonal density matrices $\f{\rho}(t)$ remain diagonal at time $t+\tau$, which justifies our second main assumption in the main text.

From the above considerations, we can also calculate the change of the system energy
\begin{align}
\Delta E_S &= E_S(t+\tau)-E_S(t)\nn
&= \int\limits_t^{t+\tau} dt_1 dt_2 \sum_{\alpha\beta} C_{\alpha\beta}(t_1-t_2)\times\nn
&\quad\times
{\rm Tr}\Big\{[\f{S_\alpha}(t_1) H_S \f{S_\beta}(t_2)\nn
&\qquad-\Theta(t_1-t_2) H_S \f{S_\alpha}(t_1) \f{S_\beta}(t_2)\nn
&\qquad-\Theta(t_2-t_1) \f{S_\alpha}(t_1) \f{S_\beta}(t_2) H_S]\f{\rho_S}(t)\Big\}\,.
\end{align}
Representing the system coupling operators in the energy eigenbasis $\f{S_\alpha}(t) = \sum_{ab} S_\alpha^{ab} \ket{a}\bra{b} e^{\ii(E_a-E_b)t}$ and assuming that $\f{\rho}(t) = \sum_a \rho_S^{aa}(t) \ket{a}\bra{a}$ is initially diagonal in that energy eigenbasis, we eventually obtain the expression
\begin{align}\label{EQ:rates_mic}
\Delta E_S &=  \sum_{ab} \sum_{\alpha\beta} \int\limits_0^{\tau} dt_1 dt_2 C_{\alpha\beta}(t_1-t_2) S_\alpha^{ab} S_\beta^{ba}\times\nn
&\qquad\times e^{\ii(E_a-E_b)(t_1-t_2)} (E_b-E_a) \rho_S^{aa}(t)\nn
&\equiv \sum_{ab} \tau  (E_b-E_a) R_{a\to b}^\tau \rho_S^{aa}(t)\,,\nn
R_{a\to b}^\tau &= \sum_{\alpha\beta} S_\alpha^{ab} S_\beta^{ba} \int d\omega \gamma_{\alpha\beta}(\omega)\times\nn
&\qquad\times \frac{\tau}{2\pi} {\rm sinc}^2 \left[\frac{(E_a-E_b-\omega)\tau}{2}\right]\nn
&\equiv \int \tilde R_{a\to b}^\tau(\omega) d\omega\,,
\end{align}
where we have inserted the Fourier transform $\gamma_{\alpha\beta}(\omega) = \int C_{\alpha\beta}(\tau) e^{+\ii\omega\tau} d\tau$ and the effective transition rate $R_{a\to b}^\tau$  -- equivalent to the coarse-graining approach with coarse-graining time $\tau$~\cite{schaller2008a}.
For large $\tau\to\infty$ this approaches the Fermi golden rule limit $R_{a\to b}^\infty = \sum_{\alpha\beta} S_\alpha^{ab} S_\beta^{ba} \gamma_{\alpha\beta}(E_a-E_b)$ that satisfies detailed balance $R_{b\to a}^\infty = e^{-\beta (E_a-E_b)} R_{a\to b}^\infty$ and for short coarse-graining times the rates scale linearly in $\tau$.

Under analogous assumptions we can also calculate the evolution of the bath energy
\begin{align}
\Delta E_B &= E_B(t+\tau)-E_B(t)\nn
&= \sum_{\alpha\beta} \int\limits_t^{t+\tau} dt_1 dt_2 \Big[\traceS{\f{S_\alpha}(t_1) \f{\rho_S}(t) \f{S_\beta}(t_2)}\times\nn 
&\qquad\times\traceB{H_B \f{B_\alpha}(t_1) \bar\rho_B \f{B_\beta}(t_2)}\nn
&\quad-\Theta(t_1-t_2) \traceS{\f{S_\alpha}(t_1) \f{S_\beta}(t_2) \f{\rho_S}(t)}\times\nn
&\qquad\times \traceB{H_B \f{B_\alpha}(t_1) \f{B_\beta}(t_2)\bar\rho_B}\nn
&\quad-\Theta(t_2-t_1) \traceS{\f{\rho_S}(t) \f{S_\alpha}(t_1) \f{S_\beta}(t_2)}\times\nn
&\qquad\times \traceB{H_B \bar\rho_B\f{B_\alpha}(t_1) \f{B_\beta}(t_2)}\Big]\nn
&= \sum_{\alpha\beta} \int\limits_t^{t+\tau} dt_1 dt_2 \traceS{\f{S_\alpha}(t_1) \f{S_\beta}(t_2) \f{\rho_S}(t)}\times\nn
&\qquad\times \traceB{\left[\f{B_\alpha}(t_1), H_B\right] \f{B_\beta}(t_2) \bar\rho_B}\nn
&= \sum_{ab}\sum_{\alpha\beta} \int\limits_t^{t+\tau} dt_1 dt_2 S_\alpha^{ab} S_\beta^{ba} e^{\ii (E_a-E_b)(t_1-t_2)} \rho_S^{aa}(t)\times\nn
&\qquad\times \ii \frac{d}{dt_1} C_{\alpha\beta}(t_1-t_2)\nn
&= \sum_{ab} \tau  \frac{\int \omega \tilde R_{a\to b}^\tau(\omega) d\omega}{R_{a\to b}^\tau} R_{a\to b}^\tau \rho_S^{aa}(t)\,.
\end{align}
Here, the rate $\tilde R_{a\to b}^\tau(\omega)$ could be interpreted as a conditional probability density of a reservoir energy change of $+\omega$ whilst the system undergoes an energetic change $E_b-E_a$.
We emphasize that for finite $\tau$ one has in general $\Delta E_B \neq - \Delta E_S$.
Only for infinitely long dissipative strokes one recovers this equivalence as $\int \omega \tilde R_{a\to b}^\infty(\omega) d\omega = - (E_b-E_a)R_{a\to b}^\infty$.
Rather, $\Delta E_S + \Delta E_B$ should be regarded as control work that always has to be invested to keep the cycle running, such that this contribution should be considered when analyzing the efficiencies or coefficient of performance of finite-time cycles.
Furthermore, the energy-resolved rates satisfy the relation $\tilde R_{b\to a}^\tau(-\omega) = e^{-\beta\omega} \tilde R_{a\to b}^\tau(+\omega)$. 

Alternatively, we can use methods of Full Counting Statistics~\cite{esposito2009a} to define a moment-generating function for the change of the reservoir energy
\begin{align}
M(\chi,\tau) &= \trace{\f{U}_{\chi/2}(\tau) \rho_S(t) \otimes \bar\rho_B \f{U}_{-\chi/2}^\dagger(\tau)}\,,\nn
\frac{d}{dt} \f{U}_{\chi/2}(t) &= -\ii e^{+\ii \chi/2 H_B} \f{H_I}(t) e^{-\ii \chi/2 H_B} \f{U}_{\chi/2}(t)\,,
\end{align}
from which moments of the change of the reservoir energy can be obtained by suitable derivatives
$\expval{(H_B-E_B^0)^n} = (-\ii \partial_\chi)^n M(\chi,\tau)|_{\chi=0}$.
Upon a second order expansion in the interaction we can write it in terms of a shifted correlation function\begin{align}
M(\chi,\tau) &=  1 + \sum_{\alpha\beta} \int\limits_t^{t+\tau} dt_1 dt_2 \trace{\f{S_\alpha}(t_1) \f{S_\beta}(t_2) \f{\rho_S}(t)}\times\nn
&\qquad\times \left[C_{\alpha\beta}(t_1-t_2-\chi)-C_{\alpha\beta}(t_1-t_2)\right]\,,
\end{align}
and the insertion of the Fourier transform allows under the same assumptions on $\f{\rho_S}(t)$ as before to write it as
\begin{align*}
M(\chi,\tau) = 1 + \tau \sum_{ab} \int d\omega  \tilde R_{a\to b}^\tau(\omega) \left(e^{+\ii\omega\chi}-1\right) \rho_S^{aa}(t)\,,
\end{align*}
which shows that indeed $\tilde R_{a\to b}^\tau(\omega)$ can be interpreted as a conditional probability density of energy increase $\omega$ in the reservoir whilst the system undergoes the transition $a\to b$.

\subsection{Specific Example}\label{APP:specific_example}

We can make these considerations more explicit for single-spinflip coupling operators (one for each spin $\alpha$)
\begin{align}
S_\alpha = \sigma^x_\alpha\,,\qquad
B_\alpha = \sum_k t_{k\alpha} \sigma^x_{k\alpha}
\end{align}
and a spin-local reservoir of two-level systems $H_B = \sum_\alpha \sum_k \frac{\omega_{k\alpha}}{2} \sigma^z_{k\alpha}$, 
where for a canonical equilibrium state $\bar\rho_B = \frac{e^{-\beta H_B}}{\trace{e^{-\beta H_B}}}$ we find for the correlation function
\begin{align}
C_{\alpha\beta}(\tau) &= \delta_{\alpha\beta} \int \frac{\Gamma(\omega) [1-f(\omega)]}{2\pi} e^{-\ii\omega\tau} d\omega\,,\nn
f(\omega) &= \frac{1}{e^{\beta\omega}-1}\,.
\end{align}
Above, we have analytically continued the reservoir spectral function
$\Gamma(\omega) = 2\pi \sum_k t_{k\alpha}^2 \delta(\omega-\omega_{k\alpha}) = \Gamma(-\omega)$ as an even function to the complete real axis. 
This implies that the Fourier transform of the correlation function is
$\gamma_{\alpha\beta}(\omega) = \delta_{\alpha\beta} \Gamma(\omega) [1-f(\omega)]$.
As the system energy eigenbasis is the $\sigma^z_\alpha$ eigenbasis, the rates~\eqref{EQ:rates_mic} become
\begin{align}
R_{a\to b}^\tau &= \delta_{\abs{\f{a}-\f{b}}/2,1} \int d\omega \Gamma(\omega) [1-f(\omega)]\times\nn
&\qquad\times \frac{\tau}{2\pi} {\rm sinc}^2 \left[\frac{(E_a-E_b-\omega)\tau}{2}\right]\,,
\end{align}
which -- with $f(-\omega)=1-f(+\omega)$ and $\delta_{\abs{\f{a}-\f{b}}/2,1}$ enforcing single spin flip energy differences -- is equivalent to Eq.\eqref{EQ:ratecg}.

\subsection{Generic properties of coarse-grained rates}

For $\tau\to\infty$ we recover from Eq.~\eqref{EQ:ratecg} -- by using the identity 
\begin{align}\label{EQ:sincproperty}
\lim\limits_{\tau\to\infty} \frac{\tau}{2\pi} {\rm sinc}^2 \left[\frac{(\Omega-\omega)\tau}{2}\right] = \delta(\Omega-\omega)
\end{align} 
and further $f(-\omega)=1-f(\omega)$ as well as $\Gamma(-\omega)=\Gamma(+\omega)$ -- the FGR rates~\eqref{EQ:rates_fgr} that satisfy detailed balance.
For finite $\tau$ however, detailed balance is not obeyed in general.

Nevertheless, in the wideband limit $\Gamma(\omega)\to \Gamma$ (e.g. take $\delta_\nu\to\infty$ in Eq.~\eqref{EQ:spectral_function} to obtain that limit)
it follows that
\begin{align}
\Delta R &\equiv R^{\nu,\tau}_{+\Omega}-R^{\nu,\tau}_{-\Omega}\\
&=\frac{\Gamma\tau}{2\pi} \int d\omega f(\omega)\times\nn
&\qquad\times
[{\rm sinc}^2((\omega-\Omega)\tau/2)-{\rm sinc}^2((\omega+\Omega)\tau/2)]\nn
&=\frac{\Gamma\tau}{2\pi} \int d\omega [2 f(\omega)-1]{\rm sinc}^2((\omega-\Omega)\tau/2)\nn
&= -\frac{\Gamma\tau}{2\pi} \int d\omega \tanh(\beta\omega/2) {\rm sinc}^2((\omega-\Omega)\tau/2)\,,\nonumber
\end{align}
from which from symmetry arguments one can deduce that this is negative when $\Omega>0$, i.e., energy-decreasing transitions are favored.
For highly peaked and specifically tuned spectral functions however, finite-time couplings may favor population inversion, which has been proposed for performance increases of heat engines at fast strokes~\cite{mukherjee2020a}.

To compute the actual heat transfer from the reservoir, we have to consider rather the energetic change of the latter.

%%%%%%%%%%%%%%%%%%%%%%%%%%%%%%%%%%%%%%%%%%%%%%%%%%%%%%%%%%%%%%%%%%%%%%%%%%%%%%%%
%%%%%%%%%%%%%%%%%%%%%%%%%%%%%%%%%%%%%%%%%%%%%%%%%%%%%%%%%%%%%%%%%%%%%%%%%%%%%%%%
\section{Onsager cycle}\label{APP:onsager}
%%%%%%%%%%%%%%%%%%%%%%%%%%%%%%%%%%%%%%%%%%%%%%%%%%%%%%%%%%%%%%%%%%%%%%%%%%%%%%%%
%%%%%%%%%%%%%%%%%%%%%%%%%%%%%%%%%%%%%%%%%%%%%%%%%%%%%%%%%%%%%%%%%%%%%%%%%%%%%%%%

The partition function per spin for the 2D anisotropic Ising model~\eqref{EQ:ising} is due to Onsager (cf. Eq.~(108) from Ref.~\cite{onsager1944a})
\begin{align}
\ln\frac{Z(\beta)}{2} &= \frac{1}{2\pi^2} \int\limits_0^\pi d\theta_x \int\limits_0^\pi d\theta_y \ln\Big[\cosh(2\beta J_x)\cosh(2\beta J_y)\nn
&\qquad- \sinh(2\beta J_x) \cos(\theta_x)-\sinh(2\beta J_y) \cos(\theta_y)\Big]\,.
\end{align}
Also a single-integral representation is possible, but may suffer from numerical instabilities for $J_x J_y <0$.
By acting with suitable derivatives on the integrand, we obtain integral representations of expectation values of some observables $X = \sum_{ij} \sigma^z_{ij} \sigma^z_{i+1,j}$ and $Y=\sum_{ij} \sigma^z_{ij} \sigma^z_{i,j+1}$ in thermal equilibrium
\begin{align}
\expval{X} &= \frac{1}{\beta} \partial_{J_x} \ln Z(\beta)\,,\qquad
\expval{Y} = \frac{1}{\beta} \partial_{J_y} \ln Z(\beta)\,,\nn
\expval{H} &= -\partial_\beta \ln Z(\beta) = -J_x \expval{X} - J_y \expval{Y}\,,
\end{align}
which can be eventually evaluated numerically.
With regard to the idealized (infinite relaxation time) quantum Otto cycle in the main text, this implies for the points at thermal equilibrium
\begin{align}
E_A &= \expval{H_c}_A = - J_x^c \expval{X}_A - J_y^c \expval{Y}_A\,,\nn
E_C &= \expval{H_h}_C = - J_x^h \expval{X}_C - J_y^h \expval{Y}_C\,.
\end{align}
As for the quenches we consider here, the state does not change during the transitions $A\to B$ and $C\to D$, we can also compute the other energy expectation values
\begin{align}
E_B &= \expval{H_h}_A = - J_x^h \expval{X}_A - J_y^h \expval{Y}_A\,,\nn
E_D &= \expval{H_c}_C = - J_x^c \expval{X}_C - J_y^c \expval{Y}_C\,.
\end{align}

%%%%%%%%%%%%%%%%%%%%%%%%%%%%%%%%%%%%%%%%%%%%%%%%%%%%%%%%%%%%%%%%%%%%%%%%%%%%%%%%
%%%%%%%%%%%%%%%%%%%%%%%%%%%%%%%%%%%%%%%%%%%%%%%%%%%%%%%%%%%%%%%%%%%%%%%%%%%%%%%%
\section{Computation of trajectories}\label{APP:stochprop}
%%%%%%%%%%%%%%%%%%%%%%%%%%%%%%%%%%%%%%%%%%%%%%%%%%%%%%%%%%%%%%%%%%%%%%%%%%%%%%%%
%%%%%%%%%%%%%%%%%%%%%%%%%%%%%%%%%%%%%%%%%%%%%%%%%%%%%%%%%%%%%%%%%%%%%%%%%%%%%%%%

The state $\f{\sigma}$ of a $100\times 100$ spin lattice may assume $2^{10000}$ different configurations.
The direct solution of such high-dimensional rate equations for probabilities $P_{\f{\sigma}}$ 
\begin{align}
\dot P_{\f{\sigma}} = \sum_{\f{\sigma}'} \left[R_{\f{\sigma}'\to \f{\sigma}} P_{\f{\sigma}'} - R_{\f{\sigma}\to \f{\sigma}'} P_{\f{\sigma}}\right]
\end{align}
thus easily exceeds computational capabilities even when the \new{(coarse-grained)} transition rates $R_{\f{\sigma}\to \f{\sigma}'}\ge 0$ are restricted to single bitflips.
Therefore, we resort to trajectory solutions composed of jumps between different states, that on average reproduce the probabilities.
The trajectories are constructed as follows:
Being in state $\f{\sigma}$ at time $t$ we compute a waiting time $\tau_{\f{\sigma}}$ for remaining in that state.
To obtain the proper statistics dictated by the transition rates, the waiting time is sampled from a uniformly distributed random number $r\in[0,1]$ 
\begin{align}
\tau_{\f{\sigma}} = \frac{-\ln(1-r)}{\sum\limits_{\f{\sigma}'\neq \f{\sigma}} R_{\f{\sigma}\to\f{\sigma}'}}\,.
\end{align}
After propagating time $t\to t+\tau_{\f{\sigma}}$, a jump is performed to a different state $\f{\sigma}''\neq\f{\sigma}$ with conditional probability
\begin{align}
P_{\f{\sigma}\to\f{\sigma}''}^{\rm jump}  = \frac{R_{\f{\sigma}\to\f{\sigma}''}}{\sum\limits_{\f{\sigma}'\neq \f{\sigma}} R_{\f{\sigma}\to\f{\sigma}'}}\,.
\end{align}
Compared to the direct solution of the rate equation, this yields a tremendous reduction in storage, as only the spin configuration $\f{\sigma}$ needs to be stored.
For large spin systems, already single trajectories may give good indication for the average behaviour.
Additionally, we exploit the translational invariance of the lattice to speed up the computations by applying a procedure known as $N$-fold way~\cite{kratzer2009a}:
We dynamically classify all spins on the lattice according to the configuration 
of their nearest neighbours into $3 \cdot 3 = 9$ different classes for which~\eqref{EQ:ising} leads to distinct energy changes upon single spin flips.
The sum over all possible transition rates can then be simplified by realizing that the rates for all members of such a class are identical.
Furthermore, the question of which jump to perform (i.e., which spin to flip eventually) can be broken down to which type of spinflip process (which class) should occur and then 
flipping a random representative spin of that class (locally updating its class and that of its neighbours).

\end{document}